\documentclass{aa}
\usepackage{epsfig}
\usepackage{rotating}
\usepackage{txfonts}
\usepackage{aalongtable,lscape}
\newcommand{\gesssim}{\mathrel{\hbox{\rlap{\hbox{\lower4pt\hbox{$\sim$}}}\hbox{$>$}}}}

\newcommand{\teff}{$T_{\rm eff}$}

\begin{document}

\title{Open clusters as key tracers of Galactic chemical
evolution. III. Element abundances in Berkeley~20, Berkeley~29, Collinder~261,
and Melotte~66.\thanks{Based on observations
collected at ESO telescopes under GTO programmes 71.D-0065, 072.D-0019, and
GO programme 076.D-0220}\fnmsep\thanks{
   Tables 4 to 7 are only available in electronic form at the CDS via anonymous
   ftp to {\tt cdsarc.u-strasbg.fr} (130.79.128.5) or via
   {\tt http://cdsweb.u-strasbg.fr/cgi-bin/qcat?J/A+A/???/???}} }

   \subtitle{}

   \author{P. Sestito\inst{1} \and A. Bragaglia\inst{2} \and 
   S. Randich\inst{1} \and R. Pallavicini\inst{3} \and 
   S.M. Andrievsky\inst{4,5} \and S.A. Korotin\inst{4,5}}

   \offprints{P. Sestito}

\institute{INAF-Osservatorio Astrofisico di Arcetri, Largo E.~Fermi 5,
             I-50125 Firenze, Italy\\
\email{sestito, randich@arcetri.astro.it}
\and
INAF-Osservatorio Astronomico di Bologna, Via C. Ranzani 1,
            I-40127 Bologna, Italy\\
\email{angela.bragaglia@oabo.inaf.it}
\and
INAF-Osservatorio Astronomico ``G.S. Vaiana'' di Palermo, Piazza del Parlamento 1, I-90134 Palermo, Italy\\
\email{pallavicini@inaf.it}
\and 
Department of Astronomy and Astronomical Observatory, Odessa National University, Shevchenko Park, 65014 Odessa, Ukraine  \\
\email{scan@deneb1.odessa.ua}
\and
Isaac Newton Institute of Chile, Odessa branch, Ukraine  }

\titlerunning{Abundances in old open clusters}
\date{Received Date: Accepted Date}

  \abstract
{Galactic open clusters are since long recognized as one of the best tools for
investigating  the radial distribution of iron and other metals.}
{We employed FLAMES at VLT to collect UVES spectra of bright giant  stars in a
large sample of open clusters, spanning a wide range of Galactocentric
distances, ages, and metallicities. We present here the results for four
clusters: Berkeley~20 and Berkeley~29, the two most distant clusters in the
sample;  Collinder~261, the oldest and the one with the minimum  Galactocentric
distance; Melotte~66.}
{Equivalent width analysis was carried out using the spectral code MOOG and
Kurucz model atmospheres to derive abundances of Fe, Al, Mg, Si, Ca, Ti, Cr,
Ni, Ba; non-LTE Na abundances were derived by direct line-profile fitting.}
{We obtain subsolar metallicities for the two anticenter clusters Be~20
([Fe/H]=$-$0.30, rms=0.02) and Be 29 ([Fe/H]=$-$0.31, rms=0.03), and for Mel~66
([Fe/H]=$-$0.33, rms=0.03), located in the third Galactic quadrant, while
Cr~261, located toward the Galactic center, has higher metallicity
([Fe/H]=+0.13, rms=0.05 dex). The $\alpha$-elements Si, Ca and Ti, and the
Fe-peak elements Cr and Ni are in general close to solar; the s-process element
Ba is enhanced. Non-LTE computations of Na abundances indicate solar scaled
values, suggesting that the enhancement in Na previously determined in giants
in open clusters could be due to neglected non-LTE effects.}
{Our results support the presence of a steep negative slope of the Fe radial
gradient up to about 10--11 kpc from the Galactic center, while in the outer
disk the [Fe/H] distribution seems flat. All the elemental ratios measured are
in very good agreement with those found for disk stars of similar metallicity
and no trend with Galactocentric distance seems to be present.}

\maketitle

\section{Introduction}\label{intro}

The radial metallicity gradient -- i.e. the behaviour of [Fe/H] with
Galactocentric radius ($R_{\rm gc}$) --  provides a powerful observational
constraint on models of Galactic disk formation and chemical evolution  (e.g.,
Tosi \cite{tosi96}; Twarog, Ashman, \& Anthony-Twarog \cite{twarog}; Freeman \&
Bland-Hawthorn \cite{freeman}). At the same time, the ratios of the abundances
of different elements to Fe, are fundamental for understanding the role of
stars with different masses in the Galactic chemical enrichment, and thus to
get insights on the initial  mass function (IMF) at the epoch of disk
formation.

Open clusters are widely recognized as ideal targets for the investigation of
chemical abundances and their evolution, due to their internal homogeneity in
age and composition, to the fact that their distance is in principle  easy to
measure, and to their broad coverage of ages and Galactocentric distances. For
this reason, many spectroscopic determinations of abundances in open clusters
have been carried out during the last years. Among them we cite  those based on
low resolution spectroscopy by Friel (\cite{friel95}) and Friel et
al.~(\cite{friel02}) which focused on metallicity ([Fe/H]),  and those based on
high resolution, by Friel et al. (\cite{friel03}; \cite{friel05})  Carretta et
al. (\cite{carretta04}, \cite{carretta05}), Carraro et al. (\cite{carraro04},
\cite{carraro6791}, \cite{carraro07}),  Yong, Carney, \& de Almeida
(\cite{yong05}), most of which present the analysis of several other elements
besides iron.

In spite of these observational efforts, various problems/discrepancies
persist, especially regarding the abundance gradients. For example,  
a consensus on the shape of the gradient has not been reached yet. Early works
suggested an unique negative slope (e.g., Friel \cite{friel95}; Carraro, Ng, \&
Portinari \cite{carraro98}, Friel et al.~\cite{friel02}); but Twarog
et al.~(\cite{twarog}) and Corder \& Twarog (\cite{corder})  favor a step-like 
distribution of the Fe content with Galactocentric distance. Very recent
results (e.g., Carraro et al. \cite{carraro04}, \cite{carraro07}; Yong et al.
\cite{yong05}) suggest that the gradient might become flat for $R_{\rm gc}$
larger than about 14 kpc.

In this context we have carried out two VLT/FLAMES programs, aimed at deriving
the gradient, based on the homogeneous analysis of a large sample of open
clusters. With the fiber link to UVES we observed giant stars in order to
obtain accurate element abundances based on high resolution data. The first
program was based on Guaranteed Time Observations (GTO; Pallavicini et al.
\cite{pallavic06}) and included three open clusters, while within the other one
(Randich et al.~\cite{messenger}) nine clusters were observed. The targets 
span a wide range of ages (from $\sim$1 Gyr up to $\sim$7 Gyr), metallicities
([Fe/H] from $\sim-0.5$ up to $\sim$+0.40)  and Galactocentric radii
($\sim$6--22 kpc); 

In Sestito et al. (2006 -- hereafter Paper~{\sc i}) and Bragaglia et al. (2008
-- hereafter Paper~{\sc ii}), the analysis of Fe and other elements in five of
the observed clusters (NGC~2660, NGC~2324, NGC~2477, NGC~3960, and Be~32) was
presented; in   Sestito et al. (\cite{ngc6253} -- hereafter S07) we reported on
abundances in the very metal rich cluster NGC~6253. In this paper we present
the analysis of the two far, old, anticenter clusters in the sample by Randich et
al.~(\cite{messenger}; Berkeley~20, and Berkeley~29), and of two old clusters
observed within the GTO program (Cr~261 and Mel~66; Pallavicini et al.
\cite{pallavic06}).  Be 20 and Be 29 are the most distant clusters in the
sample, while Cr 261 is the one with the lowest  $R_{\rm gc}$. For all these
clusters there are already papers based on high resolution spectra (see next
Section), but  we improve on past studies both on statistics (for most of them
only two stars were analyzed) and S/N ratio.

The paper is organized as follows: in Sect.~2 we describe the observational
samples, data reduction, and abundance analysis; the results are reported in
Sect.~3 and discussed in Sect.~4; a summary and conclusion are given in
Sect.~5.

\section{Observations and analysis}\label{lavoro}

\subsection{Sample clusters}\label{campioni}

\paragraph{Be~29 --} This is the most distant cluster in our sample (actually,
the most distant known open cluster) and one of the most metal poor ones. The
first photometric study of this cluster is the CCD $BVI$ survey by Kaluzny
(\cite{kalBe29}) who estimated an age of $\sim$4 Gyr and a distance of 10.5
kpc. Tosi et al.~(\cite{tosi04}) in a CCD $BVI$ study found,  depending on the
adopted stellar models, an age of 3.4 or 3.7 Gyr, $(m-M)_0$= 15.6 or 15.8
(implying $R_{\rm gc}$ = 21 or 22 kpc), with $E(B-V)$ = 0.13 or 0.10,  and
metallicity lower than solar (Z= 0.006 or 0.004), depending on the adopted 
stellar models. These values were confirmed by   Bragaglia, Held, \& Tosi
(\cite{bragaglia05}), who presented  radial velocity measurements for red giant
branch (RGB)  stars, and by Bragaglia \& Tosi~(\cite{bt06}) who give the
following parameters based on the Padova evolutionary tracks: age $\sim$ 3.7
Gyr, $(m-M)_{0}$=15.6, $E(B-V)$=0.12, $Z$=0.004, distance 13.05 kpc.

Spectroscopic investigations of Be~29 were presented by Carraro et
al.~(\cite{carraro04}), Yong et al.~(\cite{yong05}), and  Frinchaboy et
al.~(\cite{frincha06}). Carraro et al. and Yong et al. derived a metallicity
[Fe/H]=$-0.44$ (for two red clump stars) and  $-0.54$ (for two stars near the
RGB tip), respectively.  They also presented results for abundances of several
elements in addition to Fe. Frinchaboy et al. investigated the radial velocity
of the cluster, concluding that its properties are in agreement with the
Galactic anticenter stellar structure, also known as the Monoceros Ring.

\paragraph{Be~20 --} Also Be~20 is a very distant cluster, with a distance of
8.4 kpc from the Sun and $R_{\rm gc}$ of about 16  kpc. The first photometric
study of Be~20 is the CCD $VI$ survey  by MacMinn et
al.~(\cite{mcminn}); they estimated an age of   $\sim$5--6 Gyr, a reddening
$E(V-I)$=0.16, a metallicity [Fe/H]=$-$0.23 and $(m-M)_{V}$=15.0; their data
were used for the present paper. Recently Andreuzzi et al. (\cite{abt07},
\cite{abt08})
presented a new photometry, obtaining results in very good agreement with
MacMinn et al.~(\cite{mcminn}).

Two spectroscopic investigations of Be~20 were carried out so far.   Friel et
al.~(\cite{friel02}),  on the basis of low resolution  spectra of 6 stars,
derived an average [Fe/H] value of $-0.61\pm 0.14$ dex. Yong et
al.~(\cite{yong05}) derived [Fe/H]=$-$0.45 and $-$0.53 for two stars near the
RGB tip observed at high resolution.

\paragraph{Cr~261 --} This is one of the oldest open clusters in the Galaxy, 
with an age in the range 6--10 Gyr (but the upper value is most probably
spurious, see the discussion in Bragaglia \& Tosi \cite{bt06}). It is also one
of the few known old open clusters inside the Solar circle ($R_{\rm gc}\sim7.5$
kpc). Photometric studies of the cluster were carried out by Janes \& Phelps
(\cite{JP94}), Mazur, Krzeminski, \&  Kaluzny (\cite{mazur95}), and Gozzoli et
al. (\cite{gozzoli});  not surprisingly, given the low Galactic latitude and
the direction toward the center, they all agree on a high reddening values, 
ranging from $E(B-V)$=0.22 up to 0.34 (depending on the assumed metallicity). 
The Gozzoli et al. results have been recently revised by Bragaglia \& Tosi 
(\cite{bt06}), who  find a best fit solution for solar metallicity, distance
modulus  $(m-M)_{0}$=12.2, $E(B-V)$=0.30, and age 6 Gyr.

The metallicity of Cr~261 has been measured by means of low resolution
spectroscopy by Friel et al. (\cite{friel02}): from the analysis of 21 giants
they derived [Fe/H] = $-$0.16$\pm$0.13. In a more recent study the same group
determined the metallicity from high resolution data of four giants, obtaining
[Fe/H] = $-$0.22$\pm$0.05 (Friel et al. \cite{friel03}). A higher metal content
is reported by Carretta et al. (\cite{carretta05}), who derived [Fe/H] =
$-$0.03$\pm$0.03 (i.e. solar metallicity) and proposed as probable reasons for
the different result differences in the model atmospheres, in the  spectral
resolution and equivalent width ($EW$) measurements, and in the atomic
parameters (oscillator strength).

\paragraph{Mel~66 --} Deep CCD photometry for Mel~66 was provided by Kassis et
al.~(\cite{kassis}), who quote an age of 4$\pm$1 Gyr and $(m-M)_{0}$=13.2, with
[Fe/H]=$-$0.51 and $E(B-V)$ from 0.14 to 0.21 (from isochrone fitting).
Combining the distance modulus with the Galactic coordinates, an $R_{\rm gc}$
of about 10 kpc is obtained.

Previous studies of the metallicity of the cluster are those by Twarog et al.
(\cite{twarog95}) who derived [Fe/H]=$-$0.39 from $UBV$ photometry of TO stars,
and by Friel et al. (\cite{friel02}) who obtained  [Fe/H]=$-0.47\pm0.09$ from
low resolution spectroscopy of four cluster giants. The only high resolution
spectroscopic investigation carried out so far for Mel~66 is by Gratton \&
Contarini (\cite{gratton94}), who studied two giants and obtained
[Fe/H]$=-0.38\pm0.15$. 

The properties of the target clusters  are summarized in
Table~\ref{cluster_par}: references for ages, distance moduli and reddening are
also given; the Galactocentric radii were taken from  the collections by Friel
et al. (2002) and Friel (2006); slightly different values from other literature
sources were mentioned above. The [Fe/H] values cover the whole range found
in literaure (see above). 

\subsection{Observations and data reduction}\label{obs}

The target clusters were observed  with the multi-object instrument FLAMES
(Pasquini et al.~\cite{P00}) on VLT/UT2 (ESO, Chile). The fiber link to UVES
was used to obtain high-resolution spectra ($R=45,000$) for RGB and red clump
objects.

For all the clusters we performed observations  with the CD3 cross-disperser,
covering the wavelength range $\sim$4750--6800 \AA; for Cr~261 and Mel~66 we
also obtained spectra using the CD4 grating ($\sim$6600--10600 \AA). The
observations of Be~20 and Be~29 were carried out in service mode during the
period February-March 2006, while Cr~261 Mel~66 and were observed in May and
December  2003. A log of observations (date, UT, exposure time, grating
configuration, number of stars)  is given in Table~\ref{obslog}. The spectra
were reduced using the dedicated pipeline, and we analyzed the 1-d,
wavelength-calibrated spectra using standard  IRAF\footnote{IRAF is distributed
by the National Optical Astronomical Observatories, which  are operated by the
Association of Universities for Research in Astronomy,  under contract with the
National Science Foundation.}  packages. Radial velocities ($RV$) were derived
with the task RVIDLINES in IRAF; corrections for the contribution of telluric
lines were performed using TELLURIC and individual exposures were summed (see 
Paper~{\sc i} and {\sc ii}  for more details).

To maintain the highest homogeneity in our work we tried to select always the
same kind of stars, i.e., red clump stars.  For the present observations we
selected seven stars at the red clump of Mel~66,  five stars at the red clump
and one star near the RGB tip in Be~29, six stars at the clump and one slightly
brighter in  Cr~261, and six stars on the RGB near the clump level for Be~20
(since this cluster has no well defined clump).
Figures 1 and 2 show examples the spectra of all the stars observed, in a small
wavelength region.

We report information on the stars of the four clusters in Table~\ref{dataobs}:
we give the identification, equatorial coordinates (J2000), magnitudes (Cols. 
4--7),  number of exposures for each star, the average heliocentric radial
velocity ($RV$) and its rms,   the $S/N$ measured in the two wavelength regions
around $\sim$5600 and 6300 \AA, and a note (M stands for ``member", NM for
``non member", NM? for ``doubtful member"). The $B$, $V$, and $I$ magnitudes
come from the different sources referenced below for each cluster, the $K$
magnitudes are taken from 2MASS (Skrutskie et al.~\cite{skrutskie}).\footnote{
The Two Micron All Sky Survey is a joint project of the University of
Massachusetts and the Infrared Processing and Analysis Center/California
Institute of technology, funded by the national Aeronautics and Space
Administration and the National Science Foundation.}

For Be~20 we adopted the identification number used for the FLAMES pointings, 
based on the EIS catalogue used to select and point stars.  Only two stars of
the six observed turned out to be secure cluster members. These stars, 1201 and
1240, have $RV$ of about 78.5 km  s$^{-1}$, that agrees with the average radial
velocity found by Yong et al.~(\cite{yong05}:  +78.9,  $\sigma$=0.7
km~s$^{-1}$),  therefore we label the two objects as cluster  members;
furthermore, 1201 and 1240 are the two stars closest to the cluster center.
Star 1240 has also been observed by Yong et al.  (their star 8, as seen from
their table  2). Stars 1401 and 1716 are clearly $RV$ non members; the
remaining objects, 1519 and 1666, have $RV$ differing by about 6-7 km s$^{-1}$
from  those of  members. Given the spectral resolution of our data, we can
derive $RV$ with a precision of  better than 1 km s$^{-1}$, so we conclude that
1519 and 1666 are most probably non members of Be~20. Another possibility is
that they are long period binaries, but our data were obtained on too short a
baseline to verify this possibility. Magnitudes in the Johnson-Cousins system
were taken from our reference photometry (MacMinn  et  al.  1994) for the two
member stars. For stars 1666 and 1716 we used the $VI_{\rm C}$ values from
Andreuzzi et al. (2008), while for the remaining two objects we give in
Table~\ref{dataobs} the EIS magnitude, shifted to the MacMinn et al. and
Kassis et al. systems, respectively.

For Be~29 we selected five secure cluster members, as deduced from the $RV$s
measured by Bragaglia et al. (2005) and Carraro et al. (2004: our star 398,
their 801). Also the sixth target  (star 602), chosen among the red clump ones,
turned out to be a member, as shown by the very similar $RV$. The average $RV$
is 24.66$\pm$0.37 km s$^{-1}$. The identifications and $BVI_{\rm C}$ magnitudes
in Table~\ref{dataobs} are taken from the photometry by Tosi et
al.~(\cite{tosi04}), which is also in perfect agreement with Kaluzny
(\cite{kalBe29}).

The stars observed in Cr~261 share very similar $RV$s and therefore  all of
them are considered members (the average $RV$ is $-$25.43$\pm$1.11 km
s$^{-1}$). The ID adopted is the provisional one used in the FLAMES pointings,
coming from the catalogue based on WFI@2.2m data. Estimates of the reddening of
Cr~261 are very uncertain, with  values spanning from $E(B-V)$=0.22 up to 0.34
(see Span\`o \cite{tesi_spano}).  Recent studies  (e.g. Carretta et al.
\cite{carretta05}) seem to indicate a value close to 0.30, at least in the
central region.  We adopt here the WFI $BVI$ photometry calibrated by  L.
Prisinzano (private communication). 

As for as Mel~66, five of the observed stars have $RV$ around 21  km s$^{-1}$
(the mean $RV$ is +21.25$\pm$0.37  km s$^{-1}$); star 1865 has  $RV$=+17.85 km
s$^{-1}$ and is therefore labeled  as a doubtful member. Star 1614 rotates very
rapidly and was discarded from further analysis. For this cluster we adopted
the  $VI_{\rm C}$ photometry by Kassis et al. (\cite{kassis}).

\subsection{Abundance analysis}\label{analysis}

The method of analysis is described in Papers~{\sc i} and {\sc ii}; therefore
we provide here only a brief summary, and we refer the reader to those papers 
for more complete information on the line lists adopted, references for atomic
parameters and damping. First, we determined the solar abundances of Fe and
other elements, in order to fix the zero points of the abundance scale and to
minimize errors in the results. For the Sun we obtain $\log n(\rm{Fe}$~{\sc
i})=7.49$\pm$0.04  (standard deviation, or rms)  and $\log n(\rm{Fe}$~{\sc
ii})=7.54$\pm$0.03, adopting the following effective temperature, surface
gravity and microturbulence velocity: \teff=5779 K  $\log g$=4.44, and
$\xi$=0.8 km s$^{-1}$; solar abundances of other elements are reported in table
3 of Paper~{\sc ii}.

\subsubsection{Equivalent widths}

$EW$ analysis was carried out using a recent version (2006) of the code MOOG
(Sneden \cite{sneden})\footnote{http://verdi.as.utexas.edu/} and using model
atmospheres by Kurucz (\cite{kuru}).  Continuum tracing and normalization of
the spectra  were carried out using CONTINUUM within IRAF, dividing the spectra
in small regions (50 \AA) and visually checking the output. The $EW$
measurements were carried out with the program SPECTRE, developed by Chris
Sneden (see Fitzpatrick \& Sneden \cite{spectre}), which performs a Gaussian
fitting of the line profiles. The values are available in electronic Tables
4--7,  where the first two columns list the wavelengths and element, and the
others show  the corresponding $EW$ for each star. Continuum tracing and $EW$
determination are very critical steps in the analysis, possibly leading to
discrepancies among the  results of the various authors. These procedures are
particularly problematic for giant Population {\sc i} stars, that suffer for
line blending due to the high metal content and the cool  temperature. In the
case of the two most distant clusters Be~20 and Be~29 the determination of
continuum and of $EW$s has been more difficult and uncertain due to the non
optimal $S/N$.

\subsubsection{Stellar parameters}

Initial effective temperatures were derived from $B$, $V$, $I$, and $K$
photometry, applying the calibration by Alonso, Arribas, \& Martinez-Roger
(\cite{alonso}). For Be~29 and Cr~261 we used both $B-V$ and $V-K$ colors;  for
Mel~66 and Be~20 we used $V-K$ and $V-I$, after having converted the Cousin
colors by  Kassis et al. (\cite{kassis}) and MacMinn et al.~(\cite{mcminn}),
respectively, into the Johnson ones (required in the Alonso calibration) with
the formula: $V-I_{\rm J}$=1.285$(V-I_{\rm C})$ (Bessell 1979). The \teff~were
then optimized during the analysis, in order to eliminate possible trends of
$\log n$(Fe {\sc i}) vs. the excitation potential, i.e. to satisfy the
excitation equilibrium condition. The initial surface gravity was computed from
the photometry, with bolometric corrections derived following the prescriptions
by Alonso et al.~(\cite{alonso}), and with clump masses of 1.20 $M_{\odot}$ for
Be~20 (age $\sim$5--6 Gyr),  Be~29 and Mel~66 (ages $\sim$4 Gyr), 1.15
$M_{\odot}$ for Cr~261 (age $\sim$6--7 Gyr).  The final spectroscopic gravities
were chosen by imposing the ionization equilibrium condition,  i.e. $\log
n(\rm{Fe}$~{\sc ii})$-$$\log n(\rm{Fe}$~{\sc i})=0.05 (as found for the Sun).
The initial microturbulence velocity, $\xi$, was computed following  the
relationship by Carretta et al.~(\cite{carretta04}): $\xi=[1.5-0.13\times{\log
g}]$ km s$^{-1}$, and then optimized by zeroing the slope of $\log n$(Fe {\sc
i}) as a function of theoretically expected $EW$s for a fixed abundance. 

\subsubsection{Errors}

Final abundances are affected by random and systematic errors, whose main
sources are uncertainties in oscillator strengths and $EW$s, and uncertainties
in the stellar parameters.

Internal errors due to  uncertainties in the oscillator strengths are minimized
since our  scale is referred to the Sun (see also Paper~{\sc i}).  The effect
of random errors in $EW$s and in the atomic parameters on the derived abundance
for a single star is well represented by $\sigma_1$, the standard deviation
from the mean abundance based on the whole set of lines (see
Table~\ref{tab_Fe}). Errors due to uncertainties in \teff, $\log g$, and $\xi$
($\sigma_2$) were estimated  following the method by Carretta et
al.~(\cite{carretta04}), where  a detailed description of the procedure can be
found, while in Paper~{\sc i} we give the values found by us for our  clusters.
For Cr~261 and Mel~66 the estimated random errors are similar to those in NGC
3960 (Paper~{\sc i}) since the spectra of the two clusters have very similar
$S/N$: $\sim$20 K for the \teff, $\sim$0.10 dex for the gravity and 0.10 km
s$^{-1}$ for microturbulence. In the case of Be~20 and Be~29 these random
errors are slightly larger, due to the worse $S/N$: $\sim$50 K for the \teff,
$\sim$0.2 dex for the gravity and 0.15 km s$^{-1}$ for microturbulence. Then,
we computed the sensitivity of [Fe/H] (and in general of [X/H]) to  errors in
the individual stellar parameters, and we quadratically added the three
contributions in order to obtain $\sigma_2$ (see table 13 in Paper~{\sc i} and
table 12 in Paper~{\sc ii}).

The final error $\sigma_{\rm tot}$ for Fe is the quadratic sum of $\sigma_1$
and $\sigma_2$; for other elements ([X/Fe]), we computed a total error
$\sigma_{\rm 1tot}$ by quadratically adding the $\sigma_1$ on [Fe/H] and the
$\sigma_1$ on [X/H]. The reader can find an estimate of $\sigma_2$ also for the
abundances [X/H] in Table 12 of Paper~{\sc ii}. Note that the two stars in Be
20, and star 602 in Be 29 have  $\sigma_1$ values very similar to those of
stars RGB02 and  RGB11 in Cr 261, although the spectra of the latter ones have
a higher $S/N$. This result might be due to the higher [Fe/H] of Cr 261 stars,
for which  it is more difficult to determine the continuum position during the 
$EW$ measurement, and which suffer of stronger line blending.

Finally, we checked the presence of systematic errors related to the method of
analysis by analyzing two Hyades giants with the same technique (see Paper~{\sc
i}). We found abundances in agreement with other literature estimates and this
test allowed us to conclude that no large systematic uncertainties affect our
abundance scale. 

\section{Results}\label{results}

\subsection{Metallicity}\label{metal}

The results for the stellar parameters and Fe content in the three clusters
are summarized in Table~\ref{tab_Fe}, where we report the \teff~and $\log g$
(both photometric --  the starting values -- and spectroscopic -- the
optimized values),  the microturbulence $\xi$  (spectroscopic), [Fe/H] and
their errors ($\sigma_1$ and the total $\sigma$, see Sect.~\ref{analysis}),
and the number  of Fe~{\sc i} lines used for each star.
There is generally good agreement between the photometric and
spectroscopic temperatures. The largest discrepancies are for the two Be~20
stars for which the optimized \teff~are $\sim$200 K larger  than the initial
ones, and for star 1493 in Mel~66 and star RGB05 in Cr~261, for which the 
spectroscopic~\teff ~is about 150 K hotter than the photometric one.

As for the surface gravities, for Be~20, Cr~261 and for three of the six stars
in Mel~66 (1346, 1493 and 1785) the spectroscopic ones are lower than the
initial $\log g$ with differences of $\sim$0.3 dex (varying from 0.2 to 0.5 in
the extreme cases). The discrepancy between spectroscopic and photometric
$\log g$ was already discussed in Paper~{\sc i}, and can be ascribed to
several random  factors (internal errors, errors in distance moduli, in
reddening values,  in ages,  etc.) or by non-LTE effects and/or inadequacies
in classical (1-d) model atmospheres where important features such as spots,
granulation, activity, etc.~are neglected.
Such a discrepancy has also been encountered by other authors (e.g., Feltzing
\& Gustafsson \cite{feltzing}; Schuler et al. \cite{schuler03};  Allende
Prieto et al.~\cite{allende04}) in studies of cool metal-rich stars. In the
case of Cr~261 the main sources of errors are the large uncertainties in the
reddening (see also the discussion in Carretta et al. \cite{carretta05}).
On the other hand, for stars in Be~29 we found a very good agreement among the
two sets of gravities, with an average difference $-0.05\pm0.14$ dex.
Estimates of the effects of variations in the stellar parameters on the
derived abundances can be found in Table 13 of Paper~{\sc i}, and Table 12 of
Paper~{\sc ii}. The microturbulent velocities as optimized from the analysis
are in very good agreement with the initial ones computed using the Carretta
et al.~(\cite{carretta04}) formula for almost all the stars in the three
sample,  and for this reason we show only the final values in the table.

The average metallicities for the four clusters are listed in
Table~\ref{tab_Fe}: three of them have subsolar Fe contents, that is
[Fe/H]=$-0.30\pm0.02$ (Be~20), $-0.31\pm0.03$ (Be~29),  $-0.33\pm0.03$
(Mel~66); Cr~261 has a slightly oversolar metallicity, similar to that of the
Hyades ($+0.13\pm0.05$).
No significant scatter among the various stars in a single
cluster exists.
Note that star 1865 in Mel~66, labeled as a doubtful member for its $RV$
discrepant by more than three $\sigma$  from the average, has a metallicity in
perfect agreement with those of other stars, therefore we consider this star
as a possible member (perhaps a binary).
Figure~\ref{ferro} shows plots of  [Fe/H] vs.~effective temperature for the
clusters. Error bars ($\sigma_{\rm tot}$) are reported, while the solid  and
dashed lines represent the average metallicity and the rms, respectively. We
may appreciate the absence of any trend between temperature and metallicity;
this is especially visible in the two clusters Be~29 and Cr~261, where stars
covering a larger range of temperatures were observed.

As briefly mentioned in Sect.~\ref{campioni}, the previous (high resolution) 
spectroscopic metallicity determination for Be~20, Be~29 and Mel~66 also found
subsolar values,  although somewhat lower than ours. In particular,  Carraro
et al. (2004) and Yong et al. (2005) determined [Fe/H]=$-$0.44 and $-$0.54,
respectively, for two stars each in Be~29. The value of [Fe/H]$=-0.38$
measured  by Gratton \& Contarini (\cite{gratton94}) for Mel~66 is instead
closer to our value. For Cr~261 two previous analysis exist. Friel et al.
(\cite{friel03})  obtained a metallicity [Fe/H]=$-0.22 \pm 0.05$,
significantly lower than ours, while Carretta et al. (\cite{carretta05}) found
a solar [Fe/H].

\subsection{Abundances of other elements}\label{altri}

We derived the abundances  of the light elements Na and Al, the
$\alpha$-elements Mg, Si, Ca, and Ti, the Fe-peak elements Cr and Ni, and the
s-process element Ba.  Table~\ref{alpha} shows the results ([X/Fe]) for Si,
Ca, Ti, Cr, Ni, and Ba, for which a significant number of lines was employed
(see Paper~{\sc ii} for further details).

The ratios of $\alpha$-elements Si, Ca, and Ti to Fe  are close to solar in
the sample clusters, with a slight enhancement of Ca and Si in some cases
([X/Fe]$\sim$+0.10/+0.15). Also the Fe-peak elements are solar, with [Cr/Fe]
slightly below zero. The s-process Ba is sensitively enhanced, as already
found by our group (Paper {\sc ii}; Bragaglia et al.~\cite{bragaglia06});
abundance for this element, however, strongly depends on the adopted analysis
and indeed has been found to vary a lot between clusters  (e.g. Gratton,
Sneden, \& Carretta \cite{G04}). The [X/Fe] values as a function of [X/Fe] for
Si, Ca, Ti, Cr, Ni and Ba are shown in Figs.~\ref{abbondanzeBe20},
\ref{abbondanzeBe29}, \ref{abbondanzeCr261}, and \ref{abbondanzeMel66} for the
four clusters. No particular trends are present for abundances, nor a large
scatter, with only a few exceptions: the two members of Be~20 have Si
abundances differing by $\sim$0.2 dex; the tip star in Be~29 (1024) has Ti
abundance considerably higher than the average. The latter fact might be
related to problems in determining abundances in such a cool star.

The determination of Na, Mg, and Al abundances is affected by larger 
uncertainties with respect to the other species above,  since for these three
elements only a few spectral lines are available in the wavelength range
covered by our spectra, and moreover they are difficult to measure. Note that
spectra in the 8000 \AA~range were available only for Cr~261 and Mel~66, as
mentioned in Sect.~2. As already done in S07 and Paper {\sc ii}  for Al and
Mg, we decided to present abundances on a line-by-line basis.
The upper parts of Tables~\ref{linetolineBe20}, \ref{linetolineBe29},
\ref{linetolineCr261},  and \ref{linetolineMel66}  show the abundances
\emph{for each line} of the elements Al, Mg, and Na (for the latter one, see
next Section). A large scatter is present among the different lines and the
various stars, with some clear outliers being present. This is probably due to
non-optimum continuum positioning related to  quite low $S/N$. 

\subsection{Sodium}\label{sodio}

The abundances of Na, Mg, and Al computed by us using MOOG (LTE analysis) are
in general oversolar (this is not an exception; for a recent determination see
e.g., Jacobson et al. \cite{jfp07}). However, it is known that the Na spectral
lines are affected by non-LTE effects, that depend  on the temperature and 
evolutionary status of the star.  For example, Gratton et
al.~(\cite{gratton99}) computed non-LTE corrections (using statistical
equilibrium calculations) for F- and K-type stars over a broad range of
gravities. As shown in Paper~{\sc ii}, the corrections following the
prescription of Gratton et al. (1999) are about  $-0.02$ to $-0.10$ dex for
stars in our sample clusters. Slightly larger values are found following the
prescription of the work by Mashonkina  et al.~(\cite{masho}), who give
corrections up to $-0.15$ dex for giant stars (see also S07).

A more recent analysis has been carried out by Andrievsky et
al.~(\cite{andri07}), who determined non-LTE Na abundances for a homogeneous
sample of metal-poor stars  with a direct method: namely, they compute non-LTE
abundances by line profile fitting, and not by applying a correction to LTE
calculated abundances. Andrievsky et al.~(\cite{andri07}) use a  modified
version of the code MULTI (Carlsson et al.~\cite{carl86}; Korotin, Andrievsky,
\& Luck \cite{kor99a}; Korotin, Andrievsky, \& Kostynchuk \cite{kor99b}; see
additional details in Andrievsky et al.~\cite{andri07}).  The atomic model
that consists of 27 energy levels of Na~{\sc i} atom and the ground level of
Na~{\sc ii} ion was used in our NLTE calculations. The radiative transitions
between the first 20 levels of Na~{\sc i} and the ground level of Na~{\sc ii}
are considered.  Transitions between the other levels are used only for the
particle number conservation. Linearizing procedure includes 46 $b-b$ and 20
$b-f$ transitions. Radiative rates for 34 transitions are fixed.

As mentioned in Sect.~\ref{altri}, in the upper parts of Tables
\ref{linetolineBe20}, \ref{linetolineBe29},  \ref{linetolineCr261}, and
\ref{linetolineMel66} we show  the Na abundances for each line computed using
MOOG (LTE). In the last lines of the same Tables, we report the  non-LTE
average abundances, computed by fitting the profiles of the same Na lines used
in the $EW$ analysis.  The spectroscopic stellar parameters derived by us were
adopted, which provide good fits to the lines. We show both the [Na/H]$_{\rm
non-LTE}$ values derived from line fitting and the [Na/Fe]$_{\rm non-LTE}$
abundances: these ones were computed by using [Fe/H] derived by us for each
star, and $\log n({\rm Na})_{\odot}$=6.25 obtained by Andrievsky et al. (2007)
with the same line fitting method (non-LTE). As clearly visible, the final Na
abundances including non-LTE effects are nearly solar (or even lower), i.e. no
Na enhancement is seen. 

Figure \ref{fig_sodio} shows [Na/Fe]$_{\rm non-LTE}$ vs. [Fe/H] for the open
clusters analyzed so far by our group: the clusters of Papers {\sc i} and {\sc
ii} (NGC 3960, NGC 2660, NGC2324, NGC 2477 and Be 32) for which we adopted the
corrections by Gratton et al. (1999); NGC 6253 (S07) for which we used the
corrections by Mashonkina et al. (2000), and the clusters presented in this
paper, whose non-LTE abundances were calculated with the method of Andrievsky
et al. (2007). We can immediately appreciate the difference between the
different non-LTE  corrections, since  all the stars in the various clusters
have similar temperature and evolutionary status: the new computations produce
[Na/Fe] values lower by about 0.1--0.2 dex. However, our previous measurements
were only slightly  oversolar, with an average [Na/Fe] of about +0.1 dex.

In most studies of abundances in giant stars a Na enhancement was found (e.g.,
Friel et al. 2003; Yong et al. 2005; Carretta et al. 2005; Jacobson et al.
2007) and some discuss the possibility that it is a real feature (e.g.
Pasquini et al. \cite{pasquini04}). However, our computations seem to favor
the hypothesis that oversolar [Na/Fe] might rather be a consequence of a
non-adequate treatment of non-LTE effects for giant stars. The validity of
this hypothesis is reinforced by the fact that the enhancement is not seen in
unevolved cluster stars and in field dwarfs; see e.g. Randich et
al.~(\cite{randich06}) for M~67, where they also conclude that dwarfs and
giants have the same abundances, and Soubiran \& Girard (\cite{soubiran}) for
field stars. However, recently, Mishenina et al. (\cite{mishenina06}) studied
about 180 field red clump stars and measured non-LTE Na abundances in the same
way as in the present paper  finding a quite large dispersion (see their Fig.
~14)  and an average overabundance of about +0.1 dex; they also note that the
[Na/Fe] abundances do not behave in the same way for field dwarfs and giants.
The problem is not simple, and a dedicated study would be welcome.

\section{Discussion}\label{discussion}

In Table ~\ref{cluster_summary} we give a summary of the properties and
abundances of some key elements in the open clusters analyzed by our group
(this paper, Paper~{\sc i},~{\sc ii} and S07).   The Galactocentric radii are
from Friel et al. (2002), when present; otherwise they were taken from Friel
(1995), Friel (2006).  Ages are from Bragaglia \& Tosi (\cite{bt06}),
Bragaglia et al. (\cite{bragaglia06}), Tosi, Bragaglia, \& Cignoni
(\cite{tosi07}).
The left panel of Fig.~\ref{gradiente} shows the distribution of [Fe/H]  as a
function of the Galactocentric distance: the sample includes our data and
other clusters analyzed with high resolution spectroscopy\footnote{
Literature high resolution studies: 
Smith \& Suntzeff~(\cite{smith87}: NGC~2420),
Gratton \& Contarini~(\cite{gratton94}: NGC~2243),
Brown et al.~(\cite{brown96}: Mel~71, To~2),
Hamdani et al.~(\cite{hamdani00}: NGC~2360, NGC~2447),
Gonzalez \& Wallerstein (\cite{gonzalez00}: M11),
Bragaglia et al. (\cite{bragaglia01}: NGC~6819),
Randich et al.~(\cite{randich01}: IC~2602, IC~2391;
\cite{randich03}: NGC 188; \cite{randich06}: M67),
Sestito et al.~(\cite{sestito03}: NGC~6475),
Paulson et al.~(\cite{paulson03}: Hyades),
Schuler et al.~(\cite{schuler03}: M34; \cite{schuler04}: Pleiades),
Carretta et al.~(\cite{carretta04}: IC~4651, NGC~2506, NGC~6134), 
Carraro et al.~(\cite{carraro04}: Saurer~1; \cite{carraro07}:
Be~25, Be~73, Be~75, Rup~4, Rup~7), Ford et al.~(\cite{ford05}: Blanco~1),
Tautvaisiene et al. (\cite{taut05}: NGC~7789),		
Villanova et al.~(\cite{villanova05}: Be~22, Be~66),Yong et al.~(\cite{yong05}: Be 31, NGC 2141),
Pace et al. (in prep.: Praesepe).
The point representative of NGC~6791 (the most metal rich
cluster) is an average between the [Fe/H] values obtained by
Carraro et al.~(\cite{carraro6791}), Gratton et al.~(\cite{gratton06}),
and Origlia et al.~(\cite{origlia}).
We included in the plot only clusters older than $\sim$ 100 Myr; in the cases
of clusters in common between us and other authors, we chose our own
metallicity determination.
Most [Fe/H] values are based on the analysis of
giant stars, except those for the young clusters IC~2602, IC~2391, Pleiades,
Blanco~1, NGC~6475, M34, Hyades, and for the old clusters M67
and NGC~188.}.
The right panel of Fig.~\ref{gradiente} shows the radial [Fe/H] distribution
for open clusters observed at low resolution by Friel at al.~(\cite{friel02}).
The [Fe/H] values in the two panels of  Fig.~\ref{gradiente} are taken
directly from the literature sources: thus, whereas the low resolution data by
Friel et al. represents  a homogeneous sample,  our data and the other high
resolution ones from the literature could be on different scales. Whereas this
might introduce some spurious scatter, some general features are evident in
the figure. If one considers high resolution data, there appears to be a 
quite steep metallicity gradient up to a Galactocentric distance
$R_{\rm{gc}}\lesssim10-11$ kpc, and then a flattening at larger distances.
More in detail, using a weighted linear fit, we find a slope of
$-0.17\pm0.02$ dex kpc$^{-1}$ (with a correlation coefficient $-0.83$)
in the  disk region with $R_{\rm{gc}}\leq11$
kpc, while in the outer disk ($R_{\rm{gc}}>11$ kpc) the slope is, within the
errors, consistent with zero.
The sample by Friel et al. (2002) is instead limited to  
$R_{\rm{gc}}\lesssim16$ kpc, with only one cluster in the distance range 
$R_{\rm{gc}}\sim14-16$ kpc: these authors found a negative slope of
$-0.063\pm0.010$ dex kpc$^{-1}$ between $\sim$7 and 16 kpc, and they did not
discuss the possibility of a flattening of the gradient outside of a given
radius. Considering only the clusters at $R_{\rm{gc}}\leq11$ kpc, the [Fe/H]
gradient from low resolution data has a slope of $-0.09\pm0.02$ dex
kpc$^{-1}$. The difference between this value of the slope in the inner disk
and that found from high resolution data  within the same radius
($R_{\rm{gc}}\leq11$ kpc) is important, since even small differences in the
slope might lead to changes  in the infall rate and star formation efficiency
adopted in the models. This point will be discussed in more detail in a future
paper of the series, where we plan to compare our empirical results with
updated Galactic chemical evolution models. On the other hand, considering
only high resolution data, our sample and the one from the literature give
similar results;  in particular, the possibility of a  plateau in the Fe
gradient for clusters in the outer disk was already evidenced by Yong et
al.~(\cite{yong05}) who studied Be~20 and Be~29 (also in our sample), and two
clusters at $R_{\rm{gc}}\sim12$ kpc, Be~31 and NGC~2141, as well as by Carraro
et al.~(\cite{carraro04}, \cite{carraro07}).  Our clusters significantly
improve the statistics, confirming on a more solid basis the change in the slope of
the gradient, and in particular of a flattening for radii larger than $\sim$11
kpc; the average metallicity outside of this Galactocentric distance is
$-0.27\pm0.13$.
However, we have to remember that the "inner" slope is based on a  short
baseline (about 4 kpc),  since the OC closest to the Galactic centre has $R_{\rm
GC}$ slightly less than 7 kpc, and that the number or "outer" OCs is still
scarce. To improve the situation, we should concentrate on deriving the
metallicity of clusters   both nearer to the Galactic centre and in the outer
part of the disk, and in the transition region between the two slopes. The
growing number of  photometric studies dedicated to overlooked  OCs (e.g.,
Carraro et al. \cite{carraro06}, \cite{carraro07}, Ortolani et al. 
\cite{ortolani}) and the new catalogues of candidate clusters (e.g., Froebrich
et al.  \cite{froebrich}, Kronberger et al. \cite{kronberger})  will provide a
good database for future abundance studies.

Figure \ref{confronto_disco} shows [X/Fe] ratios vs. [Fe/H] for the open
clusters in our sample and for field disk stars. We show the $\alpha$-elements
Si, Ca, Ti and the Fe-peak elements Cr and Ni. We considered the disk star
catalogue by  Soubiran \& Girard~(\cite{soubiran}), which is a compilation and
rehomogenization of several recent literature studies, but we also included
the data by Bensby et al.~(\cite{bensby}), since the latter authors derived
also Cr abundances,  not included in the other catalogue. It is evident that
the open clusters follow the same distribution of chemical abundances with
[Fe/H] of disk stars, with average values close to (but slightly larger than)
solar. Note that the $\alpha$-elements  have a certain amount of scatter, in
particular Ti. 

Figure \ref{grad_altri} shows the radial gradients for $\alpha$-elements, 
Fe-peak elements and the s-process element Ba, based on our sample and the
other high resolution data from the literature. 
The plots indicate some
amount of dispersion at any given  Galactocentric distance (this is especially
true for Ba),  with no obvious
trends, with the only exception of  Ca: in this case there seems to be a
gradient with a positive slope (0.08$\pm$0.02, with correlation coefficient
0.64)  in the inner disk.  The interpretation of
the distributions of these elements is rather complex and it is strictly
related to the role of SNe{\sc I}a and SNe{\sc II} in the Galactic chemical
enrichment (see, e.g. Yong et al. 2005). The comparison between our data and
Galactic chemical evolution models,  planned for a future paper of the
series,  will be aimed to distinguish the main contribution  in element
production by different nucleosynthesis process. 

Whereas we find nearly solar [X/Fe] ($\alpha$ and Fe-peak elements)  in all
the clusters, Yong et al. (2005) and Carraro et al. (2004) claim that the
outer disk clusters Sau~1, Be~20 and Be~29 (the two latter in common with the
present study) have  an enhancement in the $\alpha$-elements. Note, however,
that Yong et al. (2005) derive [Si+Ca/Fe]$\sim$+0.1, i.e.  values similar to
those we find  for some stars, and which are not much higher than solar, and
have been also found in  field stars of similar metallicity ; on the other
hand, they derive [Mg+Ti/Fe]$\sim$+0.2--0.3. Also the values  for the
$\alpha$-elements found by Carraro  et al. (2004) are actually only slightly
oversolar  (average [$\alpha$/Fe] ratios of less than 0.1 for Be~29 and about
0.15  for Sau~1);  anyway, in the most recent analysis of five outer disk
clusters Carraro et al. (\cite{carraro07}) do seem to converge on solar-scaled
[$\alpha$/Fe] ratios, similar to open clusters in the solar vicinity and thin
disk stars. 
With the present data, we do not think there is a strong argument 
for a significantly increased [$\alpha$/Fe] at large Galactocentric distances.

The only s-process element analyzed by us is Ba, which is clearly enhanced for
our open clusters. At the Fe abundances  of our clusters, the Ba production is
dominated by  the s-processes in low mass stars  (see e.g. Travaglio et
al.~\cite{travaglio01} and Cescutti  et al.~\cite{cescutti06} for  a complete
discussion about this element).  The Ba enhancement, mainly in the younger
clusters, might   be explained by its high production in low mass stars,
mainly from 1 to  2 M$_{\odot}$, that were formed from $\sim$ 7 to 2 Gyr ago 
(e.g. Charbonnel et al.~\cite{charbonnel96}).  These stars contribute to the
Ba enrichment exactly in the  metallicity range covered by the examined
clusters,  while at lower metallicities the main contribution  is from
r-processes which are less effective  (see also Fig.~4 by Cescutti et
al.~\cite{cescutti06}). However, Ba abundances should be treated with
caution, since the hyperfine structure could be important, as a preliminary
analysis on available data seems to show (D'Orazi and Randich, in preparation),
contrary to what found e.g., by Mashonkina and Gehren (2001); this would
decrease the abundances. Furthermore,  Ba abundances (ours and
literature ones) show
a scatter much higher than for any other element here considered, suggesting
that  we need a careful appraisal of their reliability before attempting
meaningful interpretations.

\section{Summary}\label{summary}

This study is part of a project on
the metallicity and chemical abundances of giant
stars in open clusters, whose final goal is the determination
of radial gradients in the Milky Way based on a statistically significant sample.

We report here on the abundances of Berkeley~20, Berkeley~29,  Melotte~66,
and Collinder 261; in particular, the first two clusters are outer disk objects, and the last is
inside the solar circle, so they are critical for understanding the behavior of radial gradients.
\begin{itemize}
\item We derive subsolar Fe content ([Fe/H]$\sim-0.30$) for Be~20, Be~29 and
Mel~66, and a slightly oversolar metallicity (+0.13) for Cr~261. 
\item The abundances of $\alpha$-elements (Si, Ca, Ti) and Fe-peak elements
(Cr, Ni) are close to solar for most of the stars, with a small dispersion
from the average in certain cases. The s-process element Ba is enhanced for
all our samples, as often found for open clusters (but see the warning in
last section).
\item The abundances of Mg and Al have been derived from a few lines, which in
general suggest oversolar values with respect to Fe.
\item Na abundances were computed both using $EW$s and LTE assumptions -- in
this case  [Na/Fe] values appear to be enhanced -- and deriving  non-LTE Na
abundances using a line-profile fitting technique, finding  abundance ratios
close to solar.  This supports the idea that the claim of an enhancement in Na
for  evolved stars in open clusters might be due to neglected non-LTE effects
for spectral lines, or to a non-adequate treatment of these effects. Further
studies are needed to settle this issue.
\end{itemize}

The results for the element ratios of all the clusters in our sample have been
compared to those of disk stars, finding a very good agreement: the [X/Fe]
abundances vs. [Fe/H] of open clusters and disk stars are similar, i.e. the
distributions are flat and close to solar, with the $\alpha$-elements, Ti in
particular, showing a larger dispersion than Fe-peak elements.

We have combined our metallicities with literature ones and we find evidence
for a steep negative slope of the radial metallicity gradient up to a radius
of $\sim$10--11 kpc, while we confirm the flattening of the gradient in the
outer disk. The radial distributions of our sample show for other elements a
rather large amount of scatter without obvious trends with distance  (with
the possible exception of Ca).  The
interpretation of the distributions of $\alpha$- and Fe-peak elements  is
rather complex and it is strictly related to the role of Supernovae in the
Galactic chemical enrichment. Our observational results can be  used to
distinguish the main contributions in element production by  different
nucleosynthesis processes, such as SNe {\sc i}a and SNe {\sc ii}.  This will be
the subject of a forthcoming paper. 

\begin{acknowledgements}
P.S. acknowledges support by the Italian MIUR, under PRIN 20040228979-001. We
are grateful to L. Prisinzano and P. Span\`o for helpful discussion on cluster
photometry and to P. Montegriffo for his useful software. This research has
made use of the WEBDA database operated at the Institute for Astronomy of the
University of Vienna and of NASA's Astrophysics Data System.
\end{acknowledgements}

{}

\newpage

\begin{table*}[!] 
\caption{Target clusters and their properties.}\label{cluster_par}
\begin{tabular}{lrrlcllllll}
\hline
\hline
Cluster &  l   &   b    & Age    & [Fe/H]    &  $R_{\rm{gc}}$ & $(m-M)_{0}$ &$E(B-V)$ & $E(V-I)$  &Reference\\
        &      &        & (Gyr)  &  (lit.)   & (kpc)	      &(mag)	    & (mag)   & (mag)	  & (age, dist.mod., reddening)\\
\hline
Be~20   &203.5 &$-$17.4 &  5--6  &$-$0.5$\rightarrow$$-$0.2 & 16.4 & 14.6 &  --        & 0.16  & MacMinn et al. (1994)\\ 
Be~29   &198.0 &    8.0 &  3--4  &$-$0.7$\rightarrow$$-$0.4 & 22.0 & 15.6 & 0.12       & --    & Bragaglia \& Tosi (2006)\\ 
Mel~66  &260.5 &$-$14.2 &  3--4  &$-$0.5$\rightarrow$$-$0.4 & 10.2 & 13.2 & 0.14--0.21 & --    & Kassis et al. (1997) \\
Cr~261  &301.7 & $-$5.5 &  6--7	 &$-$0.2$\rightarrow$ ~ 0.0 &  7.5 & 12.2 & 0.30       & --    & Bragaglia \& Tosi (2006)\\
\hline
\hline
\end{tabular}
\end{table*}

\newpage

\begin{table*}
\caption{Log of the observations.}\label{obslog}
\begin{tabular}{lccccc}
\hline\hline
Cluster &Date       &UT $_{\rm{beginning}}$ &Exptime  &CD & No of \\
        &yyyy-mm-dd &hh:mm:ss               &(s)      &   & stars\\
\hline\hline
Be~20   &2006-02-09 &03:07:30  &2775	& CD3 & 6 \\
        &2006-02-11 &01:19:26  &2775	& CD3 & 6 \\
        &2006-02-11 &02:28:10  &2775	& CD3 & 6 \\
        &2006-02-11 &03:23:46  &2775	& CD3 & 6 \\
        &2006-02-11 &04:21:48  &2775	& CD3 & 6 \\
        &2006-02-12 &01:46:45  &2775	& CD3 & 6 \\
Be~29   &2006-02-12 &02:45:26  &2775	& CD3 & 6 \\
        &2006-02-12 &03:54:09  &2775	& CD3 & 6 \\
        &2006-02-13 &00:36:51  &2775	& CD3 & 6 \\
        &2006-02-13 &01:28:16  &2775	& CD3 & 6 \\
        &2006-02-13 &02:45:03  &2775	& CD3 & 6 \\
        &2006-02-13 &03:35:43  &2775	& CD3 & 6 \\
        &2006-02-18 &02:29:11  &2775	& CD3 & 6 \\
        &2006-03-06 &00:32:12  &2775	& CD3 & 6 \\
Mel~66  &2003-12-01 &06:27:38  &3000	& CD3 & 7 \\ 
        &2003-12-01 &07:24:40  &3359	& CD3 & 7 \\ 
        &2003-12-01 &08:22:14  &2188	& CD3 & 7 \\ 
        &2003-12-02 &05:25:40  &3600	& CD4 & 7 \\ 
        &2003-12-02 &06:28:43  &3600	& CD4 & 7 \\ 
        &2003-12-02 &07:29:46  &3600	& CD4 & 7 \\  
Cr~261  &2003-05-27 &23:11:47  &1200	& CD3 & 7 \\
        &2003-05-27 &23:34:38  &3600	& CD3 & 7 \\
        &2003-05-28 &00:37:16  &3600	& CD3 & 7 \\
        &2003-05-28 &01:39:51  &4090	& CD3 & 7 \\
        &2003-05-28 &03:02:50  &3600	& CD4 & 7 \\
        &2003-05-28 &04:06:38  &3600	& CD4 & 7 \\

\hline\hline
\end{tabular}
\end{table*}

\newpage

\begin{table*}[!] \footnotesize
\caption{Stars observed in the four clusters; see text  for
details.}\label{dataobs}
\begin{tabular}{llllllllllclll}
\hline
\hline
Star         &   RA	&     DEC    & $B$  &$V$ & $I$ & $K$ & no. exp. & $RV$ (rms)   & $S/N$&Notes \\
ID           &          &            &      &    &    &     &CD3  &(km s$^{-1}$)&  & \\
\hline
\multicolumn{11}{c}{Berkeley~20}  \\
OC03-001201& 05 32 36.774& +00 11 04.84 &    &16.177 &  14.929 &13.276 &6 & 78.47 (0.62) &35--60 &  M  \\
OC03-001240& 05 32 38.963& +00 11 20.37 &    &15.154 &  13.727 &11.850 &6 &78.58  (0.22) &40--80 &  M  \\
OC03-001403& 05 32  5.609& +00 12 44.04 &    &15.88  &  14.53  &12.974 &6 &$-$2.95(0.90) &40--80 &  NM \\
OC03-001519& 05 32 54.960& +00 14 07.30 &    &16.32  &  15.09  &13.653 &6 & 72.18 (0.58) &35--70 &  NM?\\
OC03-001666& 05 32 46.417& +00 15 52.19 &    &15.919 &  14.673 &13.153 &6 & 85.45 (0.51) &30--50 &  NM?\\
OC03-001716& 05 32 50.057& +00 16 16.43 &    &15.447 &  13.992 &12.233 &6 & 37.67 (0.22) &40--80 &  NM \\
\hline
\multicolumn{11}{c}{Berkeley~29}  \\
 159  & 06 53 01.602 & +16 56 21.11 &17.625& 16.627&15.574 &  14.266 &8 & 24.64 (0.85) & 35--50  & M \\
 257  & 06 53 04.320 & +16 55 39.37 &17.578& 16.608&15.548 &  13.508 &8 & 24.38 (1.21) & 35--50  & M \\
 398  & 06 53 08.081 & +16 55 40.27 &17.535& 16.591&15.524 &  14.227 &8 & 24.24 (1.16) & 25--50  & M \\
 602  & 06 52 55.488 & +16 57 39.30 &17.573& 16.579&15.469 &  14.119 &8 & 24.84 (0.57) & 25--50  & M \\
 933  & 06 53 04.486 & +16 57 44.69 &17.435& 16.447&15.322 &  13.932 &8 & 25.27 (1.03) & 40--50  & M \\
1024  & 06 53 07.133 & +16 57 12.67 &16.067& 14.458&12.795 &  10.600 &8 & 24.59 (0.42) &100--180 & M \\
\hline
\multicolumn{11}{c}{Melotte~66}  \\
1346  & 07 26 17.281 & $-$47 43 59.89 &&14.59&13.36 & 11.773 &3 & 21.55 (0.34) &80--110  & M \\
1493  & 07 26 34.571 & $-$47 42 47.03 &&14.68&13.39 & 11.823 &3 & 21.32 (0.37) &80--100  & M\\
1614  & 07 26 23.682 & $-$47 42 01.83 &&14.77&13.43 & 11.866 &3 &              & 80--100 & fast rotator  \\
1785  & 07 26 02.491 & $-$47 40 55.85 &&14.59&13.35 & 11.795 &3 & 21.60 (0.32) &90--110  & M\\
1865  & 07 25 57.923 & $-$47 40 23.16 &&14.51&13.29 & 11.736 &3 & 17.85 (0.50) &110--130 & NM?\\
1884  & 07 26 19.880 & $-$47 40 15.72 &&14.71&13.52 & 11.767 &3 & 21.04 (0.34) &80--100  & M\\
2218  & 07 26 26.646 & $-$47 37 55.66 &&14.48&13.29 & 11.826 &3 & 20.72 (0.52) &100--115 & M\\
\hline
\multicolumn{11}{c}{Collinder~261}  \\
RGB02 & 12 37 34.636&  $-$68 23 25.32&  14.384&  12.937&  11.345&  9.184 & 3 &$-$25.63 (0.32)&100--130 & M \\
RGB05 & 12 38 12.612&  $-$68 21 49.49&  15.480&  14.139&  12.635& 10.020 & 3 &$-$27.09 (0.16)& 70--100 & M \\
RGB06 & 12 38 07.229&  $-$68 22 30.82&  15.380&  14.011&  12.447& 10.343 & 3 &$-$24.91 (0.23)& 80--100 & M \\
RGB07 & 12 37 55.184&  $-$68 22 35.82&  15.394&  14.013&  12.436& 10.318 & 3 &$-$25.70 (0.27)& 60--100 & M \\
RGB09 & 12 38 12.338&  $-$68 20 31.46&  15.548&  14.212&  12.700& 10.684 & 3 &$-$25.96 (0.27)& 80--120 & M \\
RGB10 & 12 37 53.921&  $-$68 21 48.61&  15.675&  14.368&  12.873& 10.767 & 3 &$-$25.32 (0.07)& 60--90  & M \\
RGB11 & 12 37 59.670&  $-$68 23 49.60&  15.438&  14.145&  12.654& 10.651 & 3 &$-$23.42 (0.18)& 80--100 & M \\
\hline
\hline
\end{tabular}
\end{table*}

\newpage

\setcounter{table}{7}
\begin{table*}[!] \footnotesize
\caption{Stellar parameters and Fe abundances.}\label{tab_Fe}
\begin{tabular}{llllllllllllllr}
\hline
\hline
Star & \teff$_{\rm{,phot}}$ & $\log g_{\rm{phot}}$ & \teff$_{\rm{,spec}}$ & $\log g_{\rm{spec}}$& $\xi_{\rm{spec}}$ & [Fe/H] &$\sigma_{1}$ & $\sigma_{\rm{tot}}$ & N lines\\
     & (K)                 &                      & (K)                 &                     & (km s$^{-1}$) & &              &&\\
\hline
\multicolumn{10}{c}{Berkeley~20} \\
 1201 &4521 &2.47 &4700 &2.10 &1.35 &$-$0.28 &0.10 &0.13& 62\\
 1240 &4226 &1.97 &4400 &1.70 &1.27 &$-$0.31 &0.12 &0.14& 94\\
\hline
average  &&&&&&$-0.30$&&0.02 \\
\hline
\multicolumn{10}{c}{Berkeley~29} \\
  159 &4994 &2.51 &4994 &2.70 &1.17 &$-$0.28 &0.14 &0.16 & 78\\
  257 &4962 &2.49 &4930 &2.58 &1.18 &$-$0.36 &0.13 &0.15 & 76\\
  398 &5045 &2.51 &5020 &2.70 &1.17 &$-$0.31 &0.14 &0.16 & 76\\
  602 &4940 &2.47 &4970 &2.35 & 1.28&$-$0.34 &0.11 &0.14 & 80\\
  933 &4915 &2.40 &4930 &2.28 &1.25 &$-$0.30 &0.13 &0.15 & 78\\
 1024 &3954 &0.95 &4050 &1.03 &1.37 &$-$0.29 &0.12 &0.14 & 79\\
\hline
average  &&&&&&$-0.31$&&0.03 \\
\hline
\multicolumn{10}{c}{Melotte~66} \\
 1346 &4688 &2.43 &4750 &2.00 &1.17 &$-$0.37 &0.09 &0.11 &104\\
 1493 &4609 &2.42 &4770 &2.15 &1.20 &$-$0.35 &0.09 &0.11 &110\\
 1785 &4691 &2.43 &4770 &2.05 &1.20 &$-$0.30 &0.09 &0.11 &101\\
 1865 &4717 &2.41 &4717 &2.05 &1.24 &$-$0.34 &0.08 &0.10 & 98 \\
 1884 &4669 &2.47 &4750 &2.45 &1.23 &$-$0.30 &0.06 &0.08 &101 \\
 2218 &4814 &2.45 &4850 &2.39 &1.25 &$-$0.31 &0.07 &0.09 &103 \\
\hline
average  &&&&&&$-0.33$&&0.03 \\
\hline
\multicolumn{10}{c}{Collinder~261} \\
RGB02 &4465 &2.08 &4350 &1.70 &1.25 &  +0.12 &0.10 &0.12 &102\\
RGB05 &4450 &2.48 &4600 &2.00 &1.24 &  +0.14 &0.13 &0.15 &114\\
RGB06 &4566 &2.53 &4500 &2.30 &1.18 &  +0.16 &0.09 &0.11 &113\\
RGB07 &4546 &2.52 &4546 &2.15 &1.20 &  +0.18 &0.09 &0.11 &111\\
RGB09 &4652 &2.66 &4720 &2.05 &1.27 &  +0.04 &0.09 &0.11 &120\\
RGB10 &4652 &2.72 &4700 &2.35 &1.20 &  +0.20 &0.09 &0.11 &120\\
RGB11 &4709 &2.66 &4670 &2.15 &1.13 &  +0.09 &0.11 &0.13 &117\\
\hline
average  &&&&&&+0.13 &&0.05 \\
\hline
\hline
\end{tabular}
\end{table*}

\newpage

\setcounter{table}{8}
\begin{table*}[!] \footnotesize
\caption{[X/Fe] abundances and averages.
Errors are the quadratic sum of $\sigma_1$ on [X/H] and on [X/Fe].}\label{alpha}
\begin{tabular}{ccccccccccccccccc}
\hline\hline
 Star &[Si {\sc i}/Fe] ($\sigma$)
      &[Ca {\sc i}/Fe] ($\sigma$) 
      &[Ti {\sc i}/Fe] ($\sigma$)   
      &[Cr {\sc i}/Fe] ($\sigma$) 
      &[Ni {\sc i}/Fe] ($\sigma$) 
      &[Ba {\sc ii}/Fe] ($\sigma$) \\
\hline
\multicolumn{7}{c}{Berkeley~20} \\
 1201 &+  0.10 (0.11) &  +0.12 (0.17) &  +0.09 (0.16) &$-$0.10 (0.20) &  +0.07 (0.14) &  +0.07 (0.14)\\
 1240 &$-$0.09 (0.22) &  +0.03 (0.16) &  +0.16 (0.18) &$-$0.10 (0.17) &$-$0.01 (0.17) &  +0.26 (0.14) \\
\hline
Average &+0.005 (0.13) & +0.075 (0.06)&  +0.13 (0.05) &$-$0.10 (0.00) &  +0.03 (0.06) &  +0.16 (0.13) \\ 
\hline
\multicolumn{7}{c}{Berkeley~29} \\
  159 &$-$0.01 (0.22) &  +0.09 (0.18) &$-$0.01 (0.20) &$-$0.10 (0.18) &$-$0.10 (0.17) &  +0.40 (0.25) \\
  257 &  +0.13 (0.28) &  +0.12 (0.20) &  +0.02 (0.20) &$-$0.06 (0.20) &  +0.01 (0.18) &  +0.46 (0.37) \\
  398 &$-$0.01 (0.16) &  +0.14 (0.18) &  +0.07 (0.19) &$-$0.10 (0.20) &$-$0.03 (0.20) &  +0.46 (0.40) \\
  602 &  +0.05 (0.12) &  +0.04 (0.16) &  +0.06 (0.15) &$-$0.13 (0.16) &$-$0.08 (0.13) &  +0.37 (0.17) \\
  933 &$-$0.05 (0.16) &  +0.08 (0.19) &$-$0.03 (0.19) &$-$0.11 (0.18) &$-$0.04 (0.20) &  +0.37 (0.20) \\
 1024 &  +0.05 (0.14) &  +0.12 (0.15) &  +0.42 (0.22) &  +0.03 (0.29) &  +0.08 (0.21) &  +0.53 (0.12) \\
\hline
Average& +0.03 (0.06) &  +0.10 (0.04) &  +0.08 (0.16) &$-$0.08 (0.06) &$-$0.03 (0.07) &  +0.43 (0.06) \\ 
       &              &               &\multicolumn{4}{l}{+0.02 (0.04) : average [Ti/Fe] computed excluding star 1024} \\
\hline
\multicolumn{7}{c}{Melotte~66}\\
 1346 &  +0.13 (0.09) &  +0.15 (0.18) &$-$0.02 (0.14) &$-$0.01 (0.15) &$-$0.02 (0.11) &  +0.32 (0.13) \\
 1493 &  +0.14 (0.10) &  +0.09 (0.15) &$-$0.02 (0.14) &  +0.02 (0.15) &$-$0.01 (0.14) &  +0.30 (0.10)  \\
 1785 &  +0.10 (0.10) &  +0.10 (0.17) &  +0.01 (0.13) &$-$0.06 (0.11) &$-$0.05 (0.12) &  +0.29 (0.10)  \\
 1865 &  +0.14 (0.11) &  +0.07 (0.14) &$-$0.05 (0.12) &$-$0.05 (0.14) &$-$0.02 (0.11) &  +0.47 (0.09)  \\
 1884 &  +0.10 (0.10  &  +0.06 (0.13) &  +0.08 (0.11) &$-$0.02 (0.13) &   0.00 (0.09) &  +0.26 (0.10) \\
 2218 &  +0.16 (0.11) &  +0.08 (0.14) &  +0.01 (0.11) &  +0.03 (0.15) &  +0.03 (0.11) &  +0.36 (0.10) \\
\hline
Average& +0.13 (0.02) &  +0.11 (0.05) &  +0.02 (0.04) &$-$0.02 (0.04) &$-$0.01 (0.03) &  +0.33 (0.07) \\
\hline
\multicolumn{7}{c}{Collinder~261}\\
RGB02 &  +0.17 (0.22) &  +0.03 (0.20) &  +0.24 (0.20) &  +0.08 (0.21) &  +0.08 (0.18) &  +0.30 (0.11)\\
RGB05 &  +0.03 (0.16) &$-$0.04 (0.20) &$-$0.03 (0.17) &$-$0.05 (0.18) &   0.00 (0.16) &  +0.26 (0.13)\\
RGB06 &  +0.14 (0.19) &$-$0.12 (0.14) &  +0.06 (0.15) &$-$0.04 (0.17) &  +0.11 (0.16) &  +0.29 (0.12)\\
RGB07 &  +0.03 (0.25) &$-$0.03 (0.15) &  +0.22 (0.19) &  +0.04 (0.19) &  +0.10 (0.17) &  +0.23 (0.10)\\
RGB09 &  +0.07 (0.15) &$-$0.03 (0.16) &  +0.07 (0.14) &  +0.10 (0.18) &  +0.06 (0.14) &  +0.31 (0.09)\\
RGB10 &  +0.09 (0.12) &$-$0.03 (0.15) &  +0.14 (0.16) &  +0.07 (0.18) &  +0.04 (0.17) &  +0.25 (0.09) \\
RGB11 &  +0.09 (0.18) &$-$0.07 (0.16) &$-$0.01 (0.15) &   0.00 (0.19) &  +0.11 (0.17) &  +0.31 (0.11)\\
\hline
Average& +0.09 (0.05) &$-$0.04 (0.05) &  +0.10 (0.10) &  +0.03 (0.06) &  +0.07 (0.04) &  +0.28 (0.03)\\
\hline
\hline
\end{tabular}
\end{table*}

\newpage

\setcounter{table}{9}
\begin{table*}[!] \footnotesize
\caption{Abundances of  Na (LTE and non-LTE), Mg, and Al for stars in Be~20.
The mean non-LTE Na abundances shown in the last line were computed from line
profile fitting.}\label{linetolineBe20}
\begin{tabular}{cccccccccccccccccccc}
\hline
\hline
\scriptsize
Be~20 \\
          &            &\multicolumn{2}{c}{1201} &\multicolumn{2}{c}{1240}\\
\hline
Wavelength &$\log n$(X)$_{\odot}$ &[X/H] &[X/Fe] &[X/H] &[X/Fe]\\
\multicolumn{6}{c}{Mg~{\sc i}}\\
  6318.71  & 7.54                 &7.51  &+0.25  & 7.62 &+0.39\\
  6319.24  & 7.47                 &--    &----   & 7.36 &+0.20\\
\multicolumn{6}{c}{Al~{\sc i}}\\
 6696.032  & 6.29                 &6.14  &+0.13  &6.31  &+0.33 \\
 6698.669  & 6.18                 &5.93  &+0.03  &6.04  &+0.17 \\
\multicolumn{6}{c}{Na~{\sc i}(LTE)}\\
  5688.22  & 6.20                 &6.25  &+0.33  &5.71  &-0.18\\
  6154.23  & 6.29                 &6.34  &+0.33  &6.20  &+0.22\\
  6160.75  & 6.30                 &6.14  &+0.12  &6.18  &+0.19\\
\hline
 & 
 & [Na/H]$_{\rm NLTE}$
 &[Na/Fe]$_{\rm NLTE}$
 & [Na/H]$_{\rm NLTE}$
 &[Na/Fe]$_{\rm NLTE}$\\
Average Na 
 & 
 &5.90 $\pm$0.10 
 &$-0.07\pm0.14$ 
 &5.90 $\pm$0.10 
 &$-0.04\pm0.16$\\
\hline
\hline
\end{tabular}
\end{table*}

\newpage

\setcounter{table}{10}
\centering
\begin{landscape}
\begin{table}[!] \scriptsize
\caption{Abundances of Na, Mg, and Al in Be~29. The mean non-LTE Na abundances
shown in the last line were computed from line profile
fitting.}\label{linetolineBe29}
\begin{tabular}{cccccccccccccccccccc}
\hline
\hline
\scriptsize
Be~29 \\
&&\multicolumn{2}{c}{159}&\multicolumn{2}{c}{257}&\multicolumn{2}{c}{398}&\multicolumn{2}{c}{602}&\multicolumn{2}{c}{933}&\multicolumn{2}{c}{1024}\\
\hline
Wavelength& $\log n$(X)$_{\odot}$&[X/H]&[X/Fe]&[X/H]&[X/Fe]&[X/H]&[X/Fe]&[X/H]&[X/Fe]&[X/H]&[X/Fe]&[X/H] &[X/Fe]\\  
\multicolumn{14}{c}{Mg~{\sc i}}\\
  6318.71   & 7.54 & 7.48 &  +0.22 & 7.00 &$-$0.18 & 7.49 &  +0.26 & 7.58 &  +0.38 & 7.51 &  +0.27 & 7.56 &  +0.31 \\
  6319.24   & 7.47 & ---- &----    & 7.43 &  +0.32 & ---- &  ----  & 7.20 &  +0.07 & 7.17 &   0.00 & 7.31 &  +0.13\\
\multicolumn{14}{c}{Al~{\sc i}}\\
 6696.032   & 6.29 & 6.03 &  +0.02 & ---- &  ----  & 6.01 &  +0.03 & 5.85 &$-$0.10 & 6.05 &  +0.06 & 6.47 &  +0.47\\
 6698.669   & 6.18 & 5.73 &$-$0.17 & ---- &  ----  & 6.05 &  +0.18 & 5.66 &$-$0.18 & 5.83 &$-$0.05 & 6.10 &  +0.21\\
\multicolumn{14}{c}{Na~{\sc i}(LTE)}\\
  5688.22   & 6.20 & 6.25 &  +0.33 & 6.14 &  +0.30 & 6.07 &  +0.18 & 6.29 &  +0.43 & 5.95 &  +0.05 & 6.25 &  +0.34 \\
  6154.23   & 6.29 & 5.48 &$-$0.53 & 5.94 &  +0.01 & 5.96 &$-$0.02 & 5.89 &$-$0.06 & 6.18 &  +0.19 & 6.24 &  +0.24 \\
  6160.75   & 6.30 & 6.03 &  +0.01 & 5.96 &  +0.02 & 6.07 &  +0.08 & 6.04 &  +0.08 & 6.12 &  +0.12 & 6.28 &  +0.27\\
\hline
        & &[Na/H]$_{\rm NLTE}$ &[Na/Fe]$_{\rm NLTE}$ &[Na/H]$_{\rm NLTE}$ &[Na/Fe]$_{\rm NLTE}$ &[Na/H]$_{\rm NLTE}$ &[Na/Fe]$_{\rm NLTE}$ &[Na/H]$_{\rm NLTE}$ &[Na/Fe]$_{\rm NLTE}$
&[Na/H]$_{\rm NLTE}$ &[Na/Fe]$_{\rm NLTE}$ &[Na/H]$_{\rm NLTE}$ &[Na/Fe]$_{\rm NLTE}$\\
Average & & 5.90$\pm$0.10         & $-0.07\pm$0.17         &5.90$\pm$0.15          &$+0.01\pm$0.20          &5.90$\pm$0.10          &$-0.04\pm$0.17          &5.90$\pm$0.10          &$0.00\pm$0.15
&5.90$\pm$0.10          &$-0.05\pm$0.16          &5.90$\pm$0.10          &$-0.01\pm$0.16\\
\hline
\hline
\end{tabular}
\end{table}
\end{landscape}

\newpage

\setcounter{table}{11}
\begin{landscape}
\begin{table}[!] \scriptsize
\caption{Abundances of Na, Mg, and Al in Mel~66. The mean non-LTE Na abundances 
shown in the last line were computed from line profile fitting.}
\label{linetolineMel66}
\begin{tabular}{cccccccccccccccccccc}
\hline
\hline
\scriptsize
Mel~66 \\
&&\multicolumn{2}{c}{1346}&\multicolumn{2}{c}{1493}&\multicolumn{2}{c}{1785}&\multicolumn{2}{c}{1865}&\multicolumn{2}{c}{1884}&\multicolumn{2}{c}{2218}\\
\hline
Wavelength& $\log n$(X)$_{\odot}$&[X/H]&[X/Fe]&[X/H]&[X/Fe]&[X/H]&[X/Fe]&[X/H]&[X/Fe]&[X/H]&[X/Fe]&[X/H]&[X/Fe]\\  
\multicolumn{14}{c}{Mg~{\sc i}}\\
  6318.71   & 7.54 & 7.45 &  +0.28 & 7.49 &  +0.30 & 7.52 &  +0.28 & 7.49 &  +0.29 & 7.49 &  +0.25 & 7.49 &  +0.26\\
  6319.24   & 7.47 & 7.51 &  +0.41 & ---  &---     & 7.50 &  +0.33 & 7.48 &  +0.35 & 7.49 &  +0.32 &---   & ---  \\
\multicolumn{14}{c}{Al~{\sc i}}\\
 6696.032   & 6.29 & 6.11 &  +0.19 & 6.01 &  +0.07 & 6.25 &  +0.26 & 6.19 &  +0.24 & 6.15 &  +0.16 & 6.20 &  +0.22  \\
 6698.669   & 6.18 & 5.98 &  +0.17 & 6.88 &  +1.05 & 6.07 &  +0.19 & 6.18 &  +0.34 & 6.24 &  +0.36 & 6.11 &  +0.23 \\
\multicolumn{14}{c}{Na~{\sc i}(LTE)}\\
  5688.22   & 6.20 & 6.08 &  +0.25 & 6.10 &  +0.40 & 6.15 &  +0.25 & 6.29 &  +0.43 & 5.99 &  +0.09 & 6.16 &  +0.27 \\
  6154.23   & 6.29 & 6.05 &  +0.13 & 6.09 &  +0.15 & 6.07 &  +0.08 & 6.31 &  +0.36 & 6.09 &  +0.10 & 6.04 &  +0.06 \\
  6160.75   & 6.30 & 6.04 &  +0.21 & 6.05 &  +0.10 & 6.11 &  +0.11 & 6.43 &  +0.47 & 6.19 &  +0.19 & 6.15 &  +0.16 \\
  8183.26   & 6.35 & 6.21 &  +0.23 & 6.20 &  +0.20 & 6.36 &  +0.31 & 6.27 &  +0.23 & 6.09 &  +0.08 & 6.29 &  +0.24 \\ 
  8194.80   & 6.35 & 6.05 &  +0.07 & 6.15 &  +0.15 & 6.18 &  +0.13 & 6.30 &  +0.26 & 6.05 &  +0.04 & 6.13 &  +0.08 \\ 
\hline
        & &[Na/H]$_{\rm NLTE}$ &[Na/Fe]$_{\rm NLTE}$ &[Na/H]$_{\rm NLTE}$ &[Na/Fe]$_{\rm NLTE}$ &[Na/H]$_{\rm NLTE}$ &[Na/Fe]$_{\rm NLTE}$ &[Na/H]$_{\rm NLTE}$
&[Na/Fe]$_{\rm NLTE}$ &[Na/H]$_{\rm NLTE}$ &[Na/Fe]$_{\rm NLTE}$ &[Na/H]$_{\rm NLTE}$ &[Na/Fe]$_{\rm NLTE}$\\
Average & & 5.95$\pm$0.10         & $+0.07\pm$0.13         &5.95$\pm$0.10          &$+0.05\pm$0.13          &5.95$\pm$0.10          &$0.00\pm$0.13           &6.07$\pm$0.10          
&$+0.16\pm$0.13          &5.93$\pm$0.10          &$-0.02\pm$0.12          &5.95$\pm$0.10          &$+0.01\pm$0.12\\
\hline
\hline
\end{tabular}
\end{table}
\end{landscape}

\setcounter{table}{12}
\begin{landscape}
\begin{table}[!] \scriptsize
\caption{Abundances of Na, Mg, and Al in Cr~261. The mean non-LTE Na abundances shown in the last line were computed from line profile fitting.}\label{linetolineCr261}
\begin{tabular}{cccccccccccccccccccc}
\hline
\hline
\scriptsize
Cr~261 \\
&&\multicolumn{2}{c}{02}&\multicolumn{2}{c}{05}&\multicolumn{2}{c}{06}&\multicolumn{2}{c}{07}\\
\hline
Wavelength& $\log n$(X)$_{\odot}$&[X/H]&[X/Fe]&[X/H]&[X/Fe]&[X/H]&[X/Fe]&[X/H]&[X/Fe]\\
\multicolumn{10}{c}{Mg~{\sc i}}\\
  6318.71   & 7.54 & 7.81 &  +0.15 & 7.90 &  +0.22 & 7.84 &  +0.14 & 7.94 &  +0.22 \\
  6319.24   & 7.47 & 7.90 &  +0.31 & 7.98 &  +0.37 & 7.92 &  +0.29 & 7.26 &$-$0.39 \\
\multicolumn{10}{c}{Al~{\sc i}}\\
6696.032    & 6.29 & 6.81 &  +0.40 & 6.60 &  +0.17 & 6.64 &  +0.19 & 6.80 &  +0.33 \\
6698.669    & 6.18 & 5.52 &  +0.22 & 6.37 &  +0.05 & 6.74 &  +0.40 & 6.68 &  +0.32 \\
\multicolumn{10}{c}{Na~{\sc i}(LTE)}\\
   5688.22  & 6.20 & 6.45 &  +0.13 & 6.31 &$-$0.03 & 6.31 &$-$0.05 & 6.43 &  +0.05 \\	
   6154.23  & 6.29 & 6.64 &  +0.23 & 6.38 &$-$0.01 & 6.56 &  +0.11 & 6.49 &  +0.02 \\
   6160.75  & 6.30 & 6.58 &  +0.16 & 6.43 &  +0.01 & 6.52 &  +0.06 & 6.64 &  +0.16 \\
   8183.26  & 6.35 & 6.57 &  +0.20 & 6.50 &  +0.01 & 6.38 &$-$0.13 & 6.46 &$-$0.07 \\
   8194.80  & 6.35 & 6.60 &  +0.23 & 6.50 &  +0.02 & 6.43 &$-$0.08 & 6.57 &  +0.04 \\
\hline
        & &[Na/H]$_{\rm NLTE}$ &[Na/Fe]$_{\rm NLTE}$ &[Na/H]$_{\rm NLTE}$ &[Na/Fe]$_{\rm NLTE}$ &[Na/H]$_{\rm NLTE}$ &[Na/Fe]$_{\rm NLTE}$ 
&[Na/H]$_{\rm NLTE}$ &[Na/Fe]$_{\rm NLTE}$\\
Average & & 6.35$\pm$0.07         & $-0.02\pm$0.12         &6.20$\pm$0.07          &$-0.19\pm$0.15          &6.30$\pm$0.07          &$-0.11\pm$0.11 
&6.38$\pm$0.07          &$-0.05\pm$0.11\\
\hline
\hline
\end{tabular}
\end{table}
\end{landscape}

\setcounter{table}{12}
\begin{landscape}
\begin{table}[!] \scriptsize
\caption{Continued -- Abundances of Na, Mg, and Al in Cr~261.}\label{linetolineCr261}
\begin{tabular}{cccccccccccccccccccc}
\hline
\hline
\scriptsize
Cr~261 \\
&&\multicolumn{2}{c}{09}&\multicolumn{2}{c}{10}&\multicolumn{2}{c}{11}\\
\hline
Wavelength& $\log n$(X)$_{\odot}$&[X/H]&[X/Fe]&[X/H]&[X/Fe]&[X/H]&[X/Fe]\\
\multicolumn{8}{c}{Mg~{\sc i}}\\
  6318.71   & 7.54 & 7.73 &  +0.15 & 7.91 &  +0.17 & 7.89 &  +0.26\\
  6319.24   & 7.47 & 7.79 &  +0.28 & 7.95 &  +0.28 & 7.94 &  +0.38\\
\multicolumn{8}{c}{Al~{\sc i}}\\
6696.032    & 6.29 & 6.62 &  +0.29 & 6.84 &  +0.35 & 6.69 &  +0.31\\
6698.669    & 6.18 & 6.36 &  +0.14 & 6.58 &  +0.20 & 6.53 &  +0.26\\
\multicolumn{8}{c}{Na~{\sc i}(LTE)}\\
   5688.22  & 6.20 & 6.42 &  +0.18 & 6.49 &  +0.09 & 6.28 &   0.00 \\
   6154.23  & 6.29 & 6.41 &  +0.07 & 6.68 &  +0.19 & 6.39 &  +0.08 \\
   6160.75  & 6.30 & 6.46 &  +0.12 & 6.57 &  +0.07 & 6.47 &$-$0.03 \\
   8183.26  & 6.35 & 6.47 &  +0.08 & 6.50 &$-$0.04 & 6.41 &   0.00 \\
   8194.80  & 6.35 & 6.45 &  +0.05 & 6.47 &$-$0.08 & 6.44 &  +0.26\\
\hline
        & &[Na/H]$_{\rm NLTE}$ &[Na/Fe]$_{\rm NLTE}$ &[Na/H]$_{\rm NLTE}$ &[Na/Fe]$_{\rm NLTE}$ &[Na/H]$_{\rm NLTE}$ &[Na/Fe]$_{\rm NLTE}$\\
Average & & 6.27$\pm$0.07         & $-0.02\pm$0.11         &6.32$\pm$0.07          & $-0.13\pm$0.11         &6.20$\pm$0.07        & $-0.14\pm$0.13\\
\hline
\hline
\end{tabular}
\end{table}
\end{landscape}

\setcounter{table}{13}
\begin{table}[!] \footnotesize
\caption{The open clusters in our sample and their main properties.}\label{cluster_summary}
\begin{tabular}{llrrrrrrrrl}
\hline
\hline
Cluster  & Age  &$R_{\rm{gc}}$ & [Fe/H]   & [Si/Fe]         & [Ca/Fe]         & [Ti/Fe]         & [Cr/Fe]         & [Ni/Fe]         & [Ba/Fe]         & Ref.\\
         & (Gyr)&(kpc)         &          &                 &                 &                 &                 &                 &                 &     \\
\hline  
NGC~6253 & 3.0  & 6.60 &   +0.36$\pm$0.07 &$-$0.01$\pm$0.05 &  +0.01$\pm$0.04 &  +0.02$\pm$0.13 &  +0.04$\pm$0.05 &  +0.05$\pm$0.05 &  +0.23$\pm$0.10 & S07 \\
Cr~261   & 6.0  & 7.52 &   +0.13$\pm$0.05 &  +0.09$\pm$0.05 &$-$0.04$\pm$0.05 &  +0.10$\pm$0.10 &  +0.03$\pm$0.06 &  +0.07$\pm$0.04 &  +0.28$\pm$0.03 & This paper \\
NGC~3960 & 0.9  & 7.96 &   +0.02$\pm$0.04 &  +0.04$\pm$0.05 &  +0.02$\pm$0.03 &$-$0.04$\pm$0.02 &  +0.05$\pm$0.03 &$-$0.01$\pm$0.03 &  +0.54$\pm$0.04 & Paper~{\sc i} \\
NGC~2477 & 1.0  & 8.94 &   +0.07$\pm$0.03 &  +0.05$\pm$0.03 &$-$0.01$\pm$0.01 &  +0.01$\pm$0.06 &  +0.02$\pm$0.03 &   0.00$\pm$0.04 &  +0.46$\pm$0.05 & Paper~{\sc ii}\\
NGC~2660 & 1.0  & 9.18 &   +0.04$\pm$0.04 &   0.00$\pm$0.03 &  +0.04$\pm$0.05 &   0.00$\pm$0.03 &  +0.00$\pm$0.02 &$-$0.03$\pm$0.02 &  +0.61$\pm$0.04 & Paper~{\sc i} \\
Mel~66   & 4.0  &10.21 & $-$0.33$\pm$0.03 &  +0.13$\pm$0.02 &  +0.11$\pm$0.05 &  +0.02$\pm$0.04 &$-$0.02$\pm$0.04 &$-$0.01$\pm$0.03 &  +0.33$\pm$0.07 & This paper \\
Be~32    & 5.5  &11.35 & $-$0.29$\pm$0.04 &  +0.12$\pm$0.04 &  +0.07$\pm$0.04 &  +0.02$\pm$0.06 &$-$0.05$\pm$0.04 &   0.00$\pm$0.04 &  +0.29$\pm$0.10 & Paper~{\sc i} \\
NGC~2324 & 0.6  &11.65 & $-$0.17$\pm$0.05 &  +0.06$\pm$0.11 &  +0.15$\pm$0.05 &$-$0.08$\pm$0.03 &$-$0.03$\pm$0.03 &$-$0.09$\pm$0.02 &  +0.62$\pm$0.08 & Paper~{\sc ii}\\
Be~20    & 6.0  &16.37 & $-$0.30$\pm$0.02 &  +0.05$\pm$0.13 &  +0.07$\pm$0.06 &  +0.13$\pm$0.05 &$-$0.10$\pm$0.00 &  +0.03$\pm$0.06 &  +0.16$\pm$0.13 & This paper \\
Be~29    & 4.0  &22.00 & $-$0.31$\pm$0.03 &  +0.03$\pm$0.06 &  +0.10$\pm$0.04 &  +0.08$\pm$0.16 &$-$0.08$\pm$0.06 &$-$0.03$\pm$0.07 &  +0.43$\pm$0.06 & This paper \\
\hline
\hline
\end{tabular}
\end{table}

\begin{figure*}
\psfig{figure=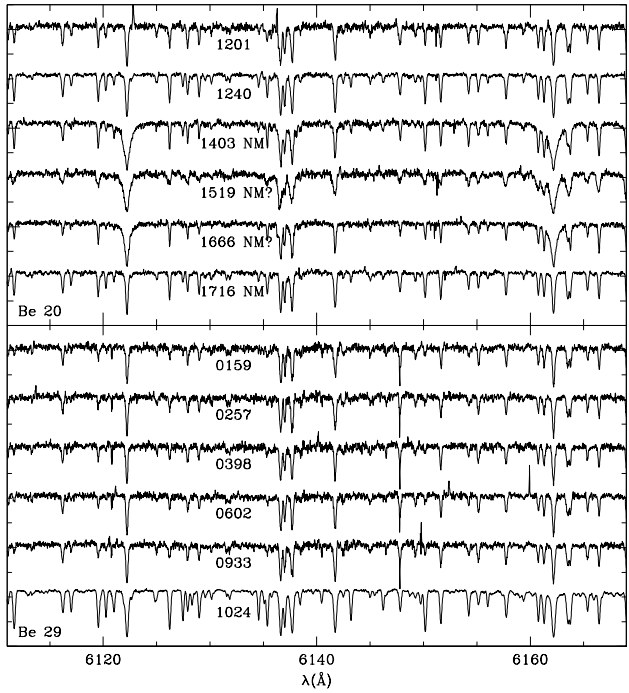,width=10cm}
\caption{A region of the spectra of stars in Be~20 and Be~29.}
\label{spe1}
\end{figure*}

\begin{figure*}
\psfig{figure=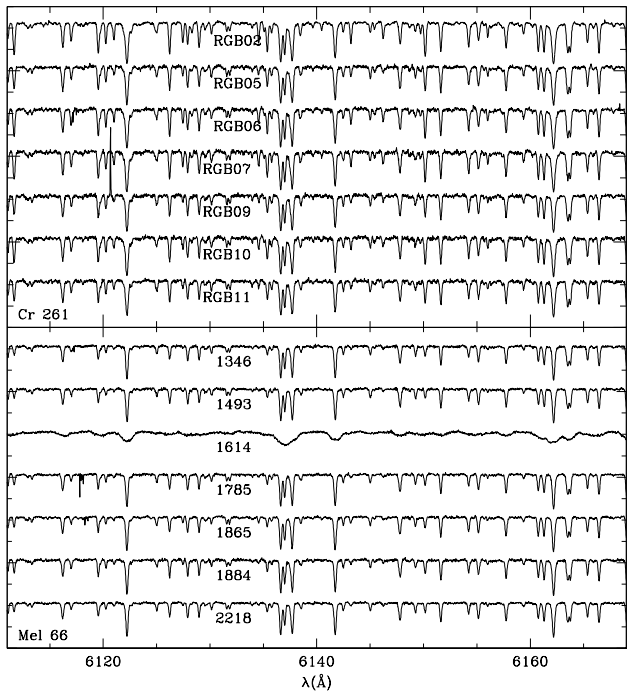,width=10cm}
\caption{A region of the spectra of stars in Cr~261 and Mel~66.}
\label{spe2}
\end{figure*}

\begin{figure*}
\psfig{figure=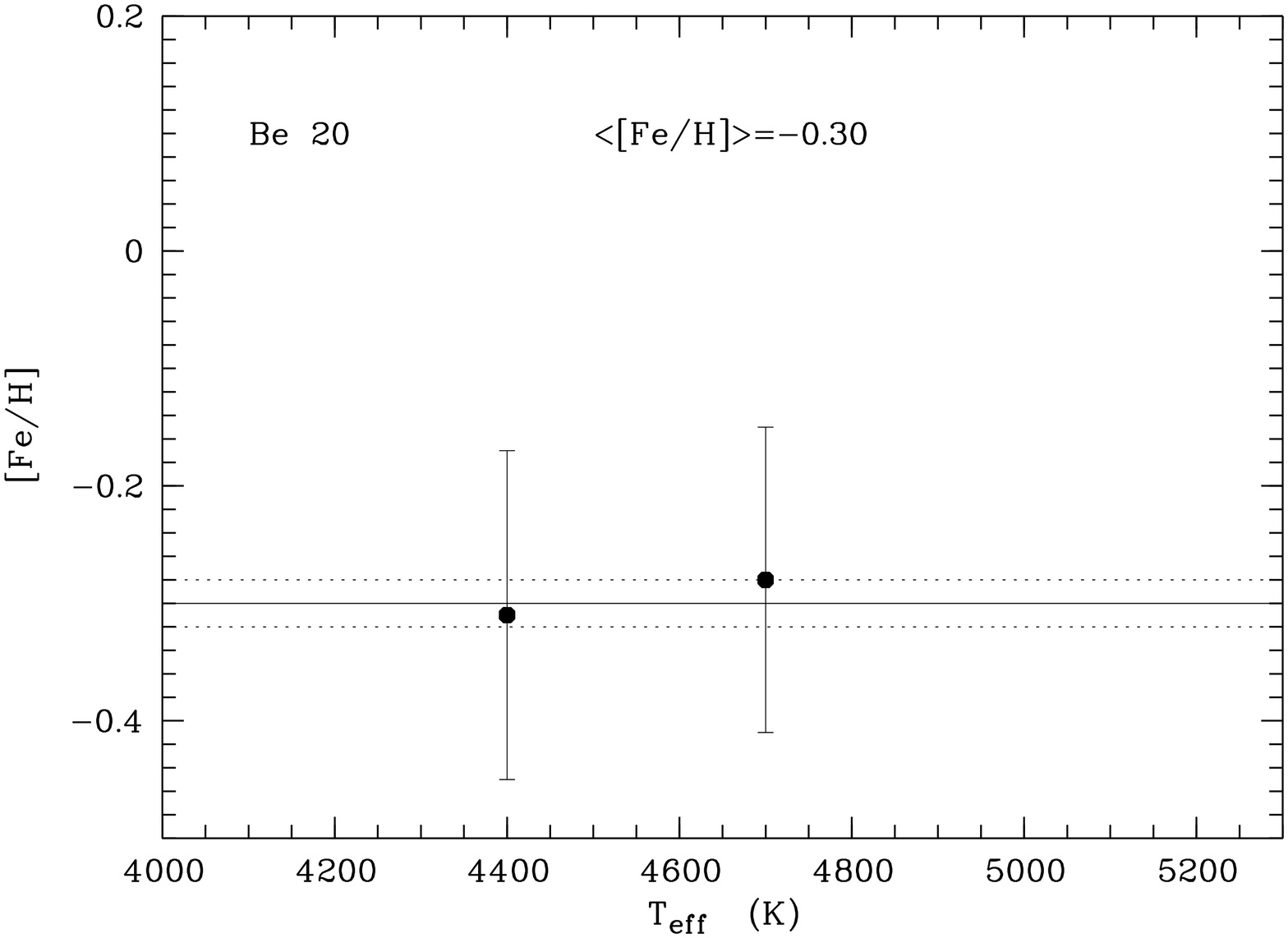, width=6cm, angle=-90}\psfig{figure=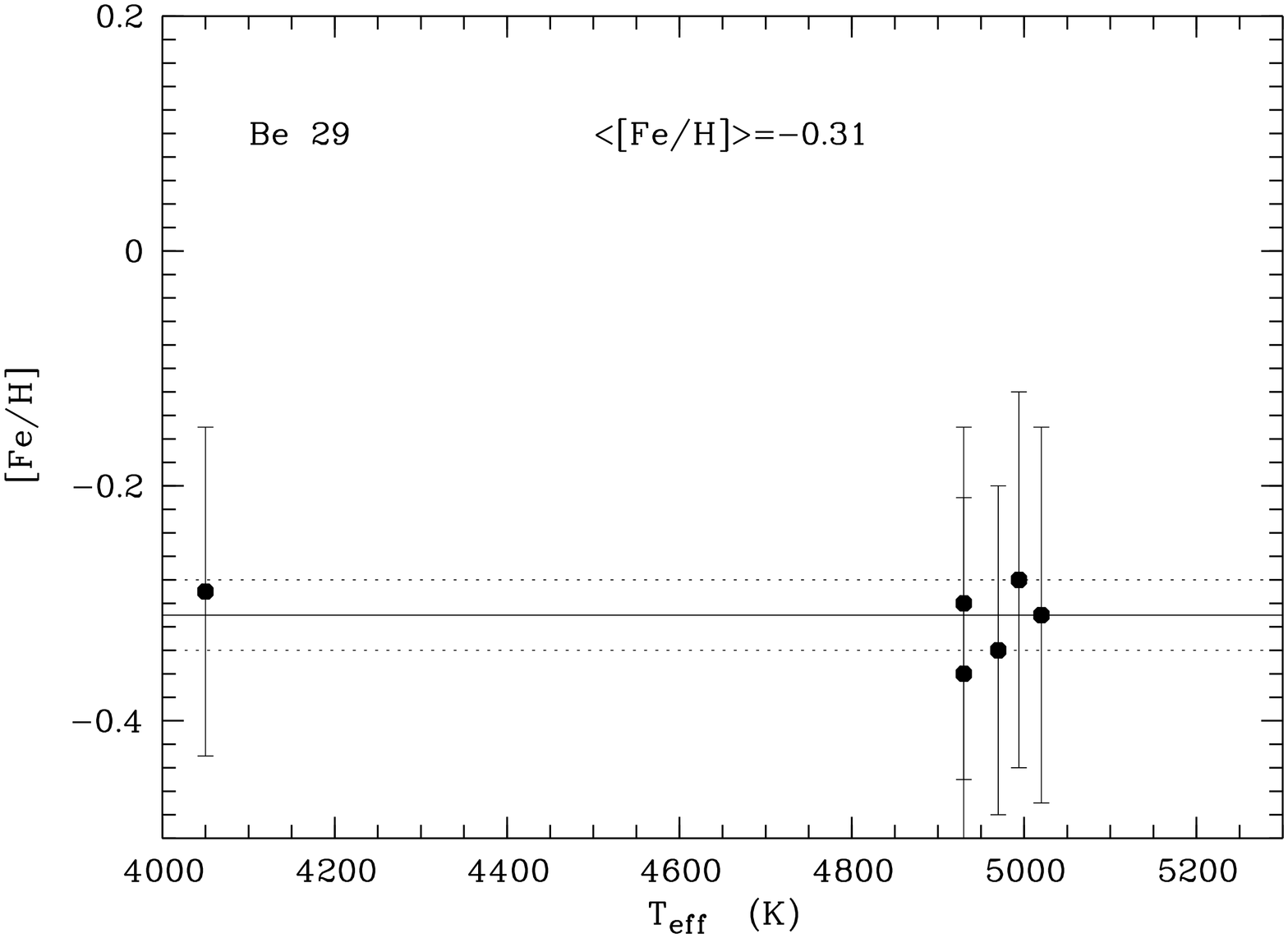, width=6cm, angle=-90}

\psfig{figure=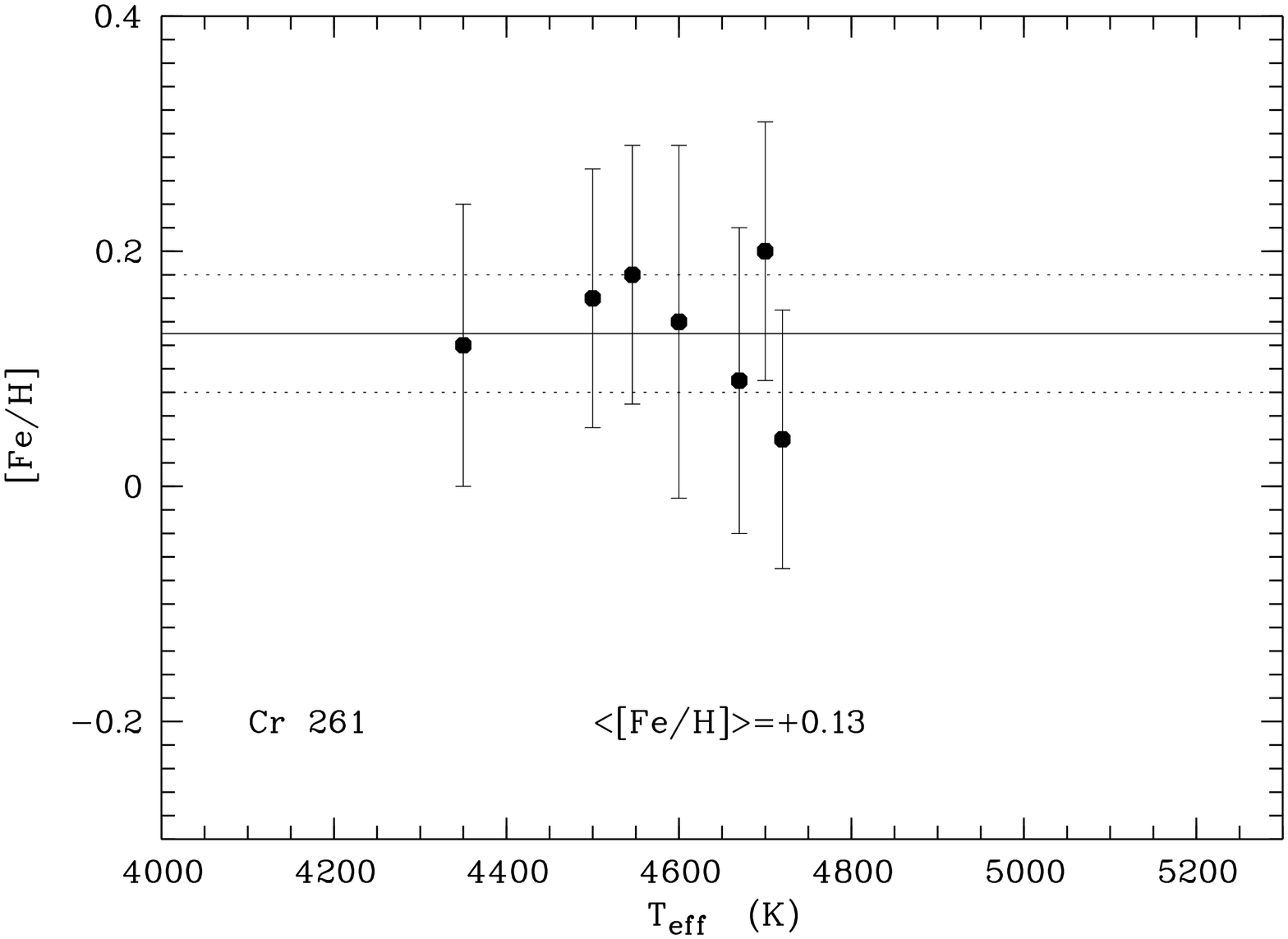, width=6cm, angle=-90}\psfig{figure=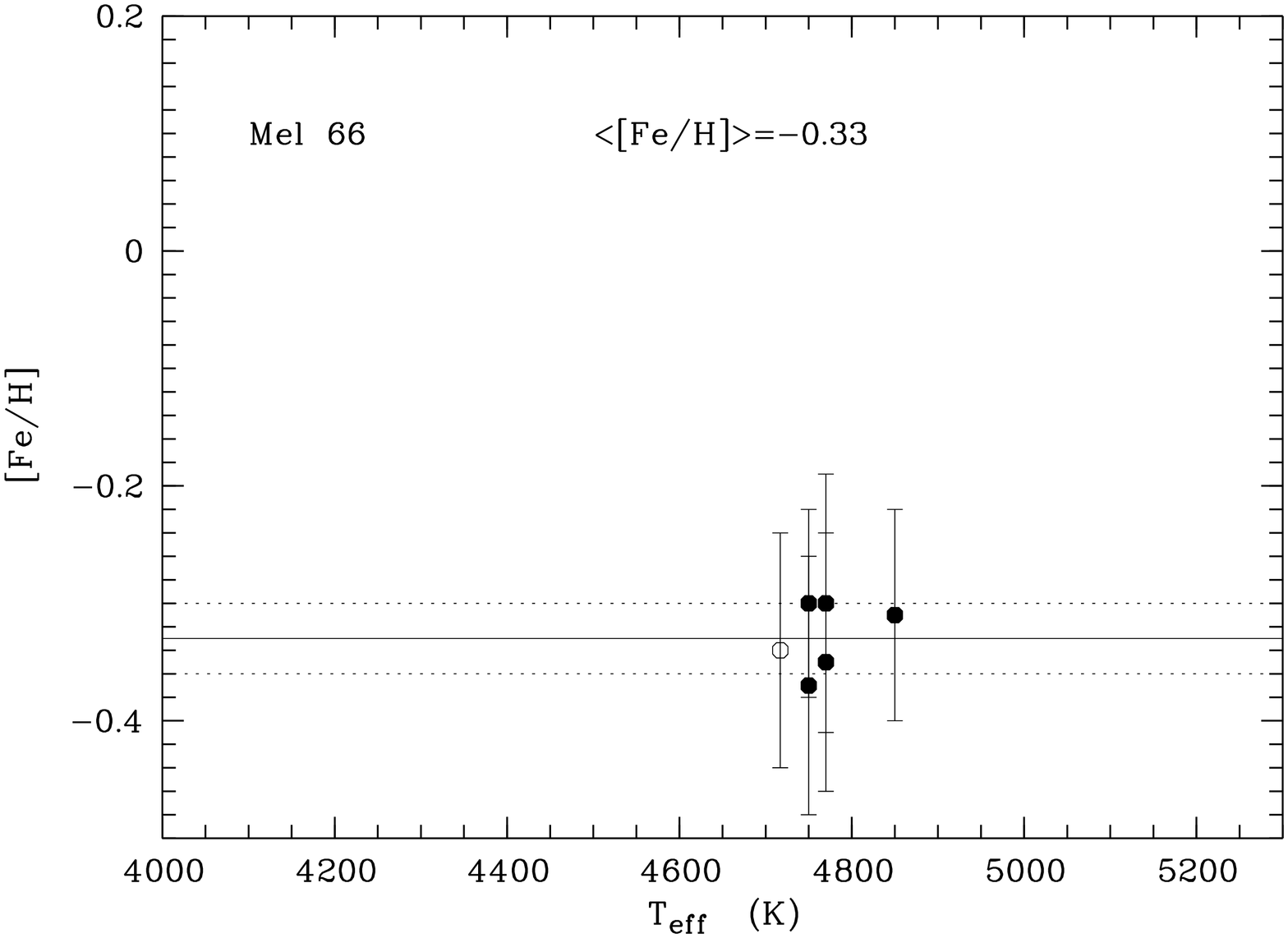, width=6cm, angle=-90}
\caption{Fe abundances as a function of \teff~for  members in the sample
clusters. The solid line represents the average Fe content, while the dashed
ones represent the rms.}\label{ferro}
\end{figure*}

\begin{figure*}
\psfig{figure=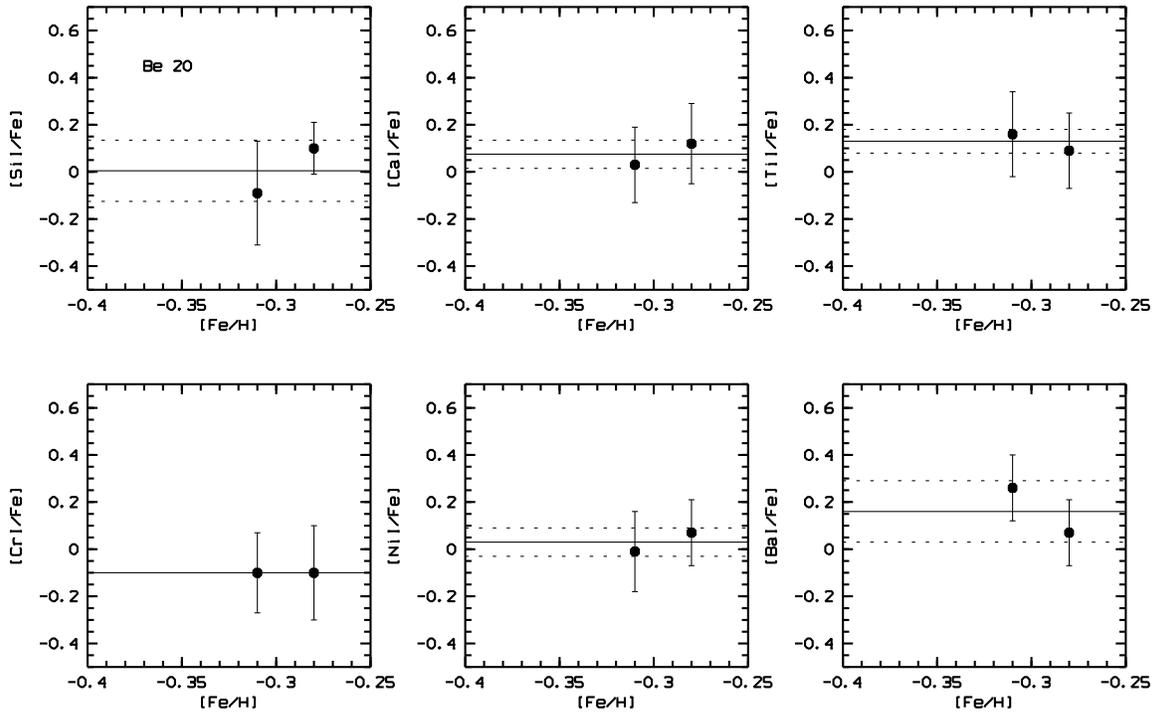, width=16cm, angle=-90}
\caption{[X/Fe] abundance ratios as a function of [Fe/H] for the two members
in Be~20.}\label{abbondanzeBe20}
\end{figure*}

\begin{figure*}
\psfig{figure=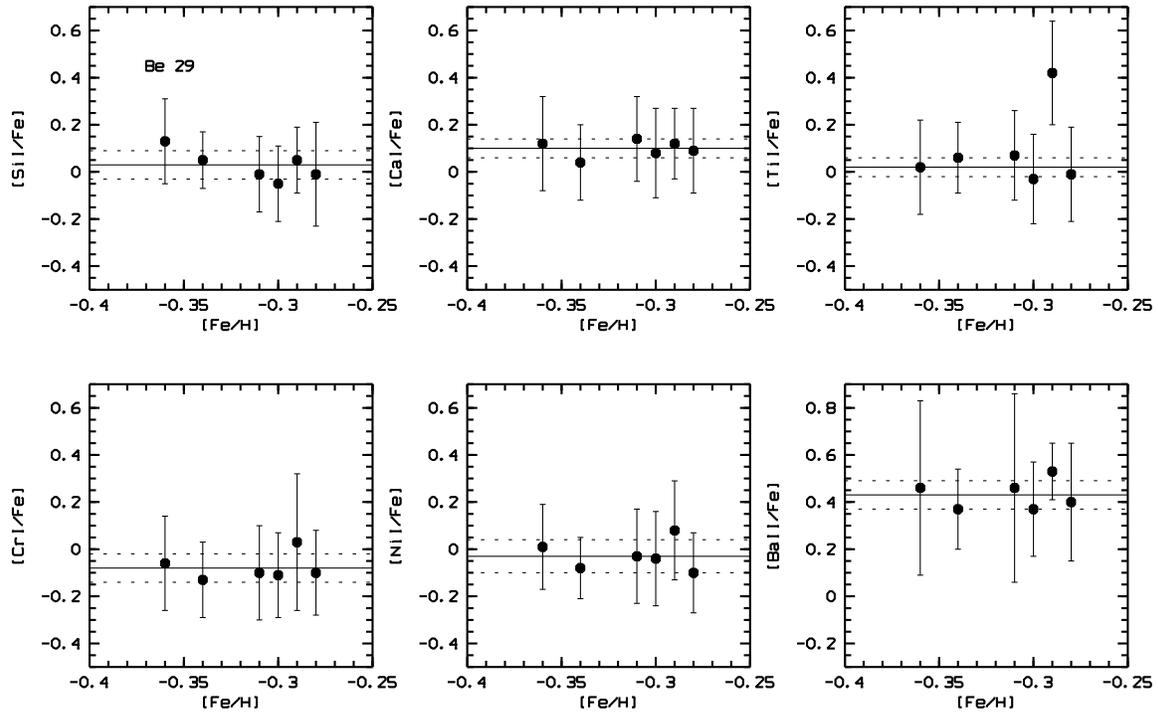, width=16cm, angle=-90}
\caption{[X/Fe] vs.~[Fe/H] for stars in Be~29. The average [Ti/Fe] has been
computed excluding the tip star 1024, since we derived a very high abundance
for this element.}\label{abbondanzeBe29}
\end{figure*}

\begin{figure*}
\psfig{figure=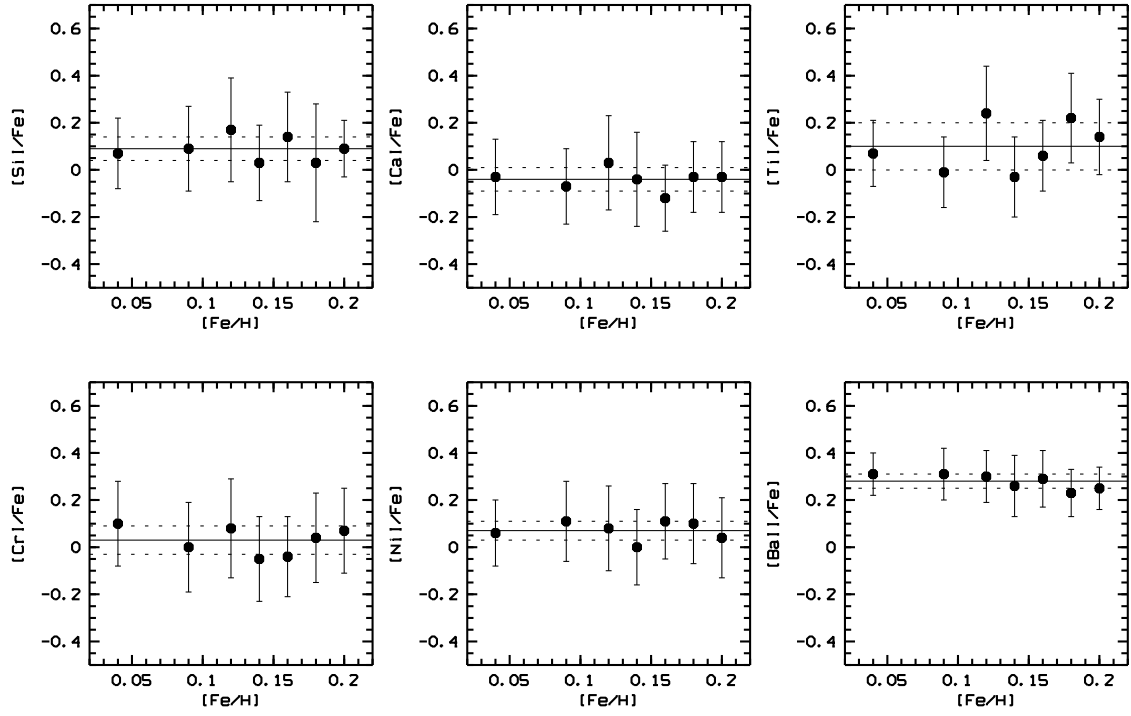, width=16cm, angle=-90}
\caption{[X/Fe] vs.~[Fe/H] for stars in Cr~261.}\label{abbondanzeCr261}
\end{figure*}

\begin{figure*}
\psfig{figure=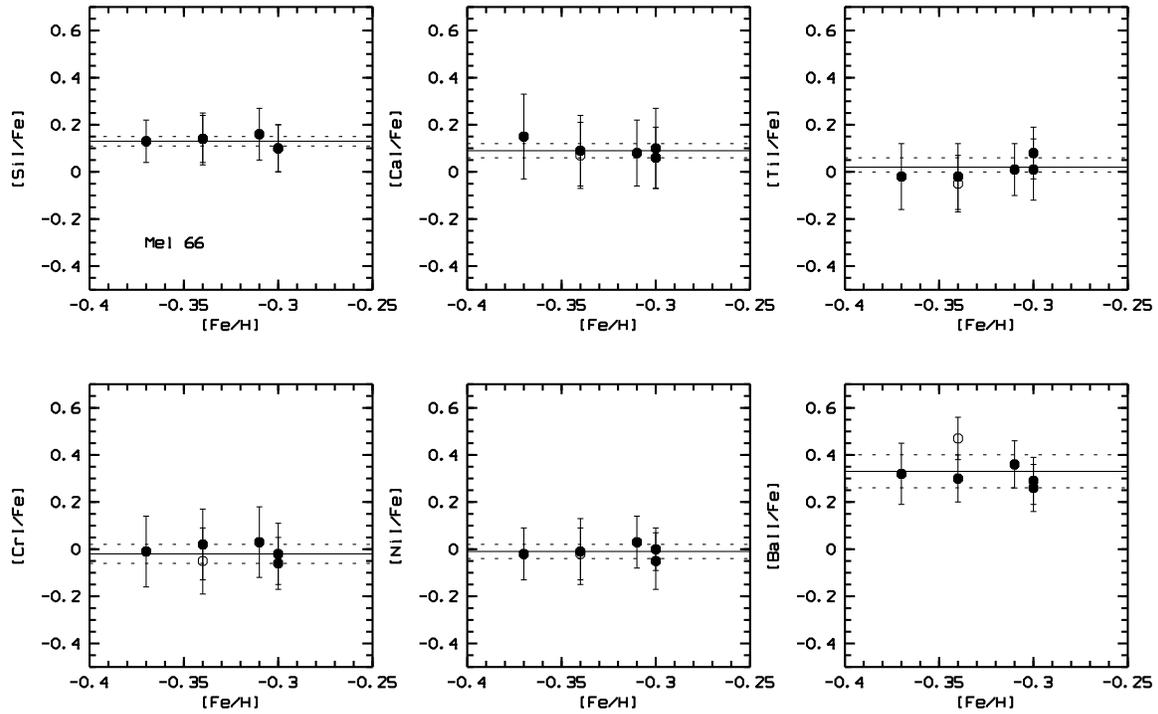, width=16cm, angle=-90}
\caption{[X/Fe] vs.~[Fe/H] for stars in Mel~66. The open symbol represents the
probable member.}\label{abbondanzeMel66}
\end{figure*}

\begin{figure*}
\psfig{figure=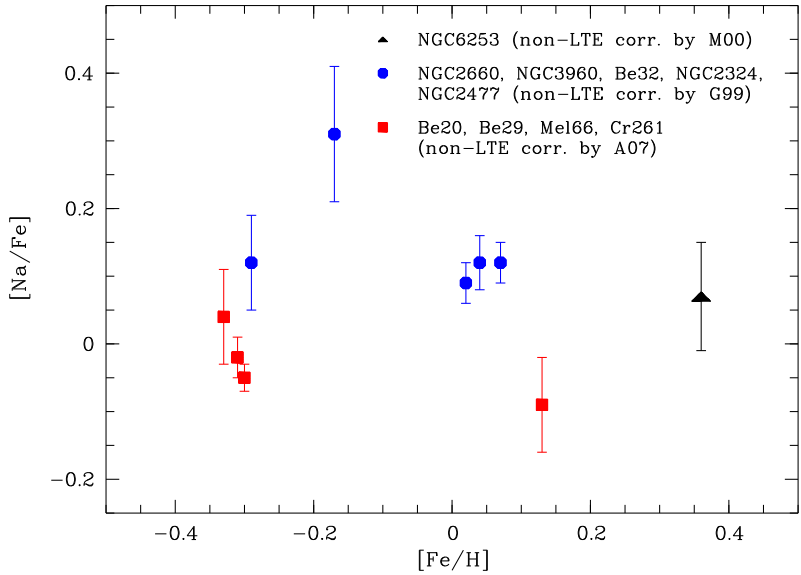, width=8cm, angle=-90}
\caption{[Na/Fe]$_{\rm non-LTE}$ vs. [Fe/H]  for clusters analyzed by us:
circles are clusters of Papers~{\sc i} and {\sc ii}, for which we adopted the
corrections by Gratton et al. (1999); the triangle is NGC 6253 (S07) for which
we used the corrections by Mashonkina et al. (2000), and squares are clusters
presented in this paper, whose abundances were computed following the profile
fitting method by Andrievsky et al. (2007).}\label{fig_sodio}
\end{figure*}

\begin{figure*}
\psfig{figure=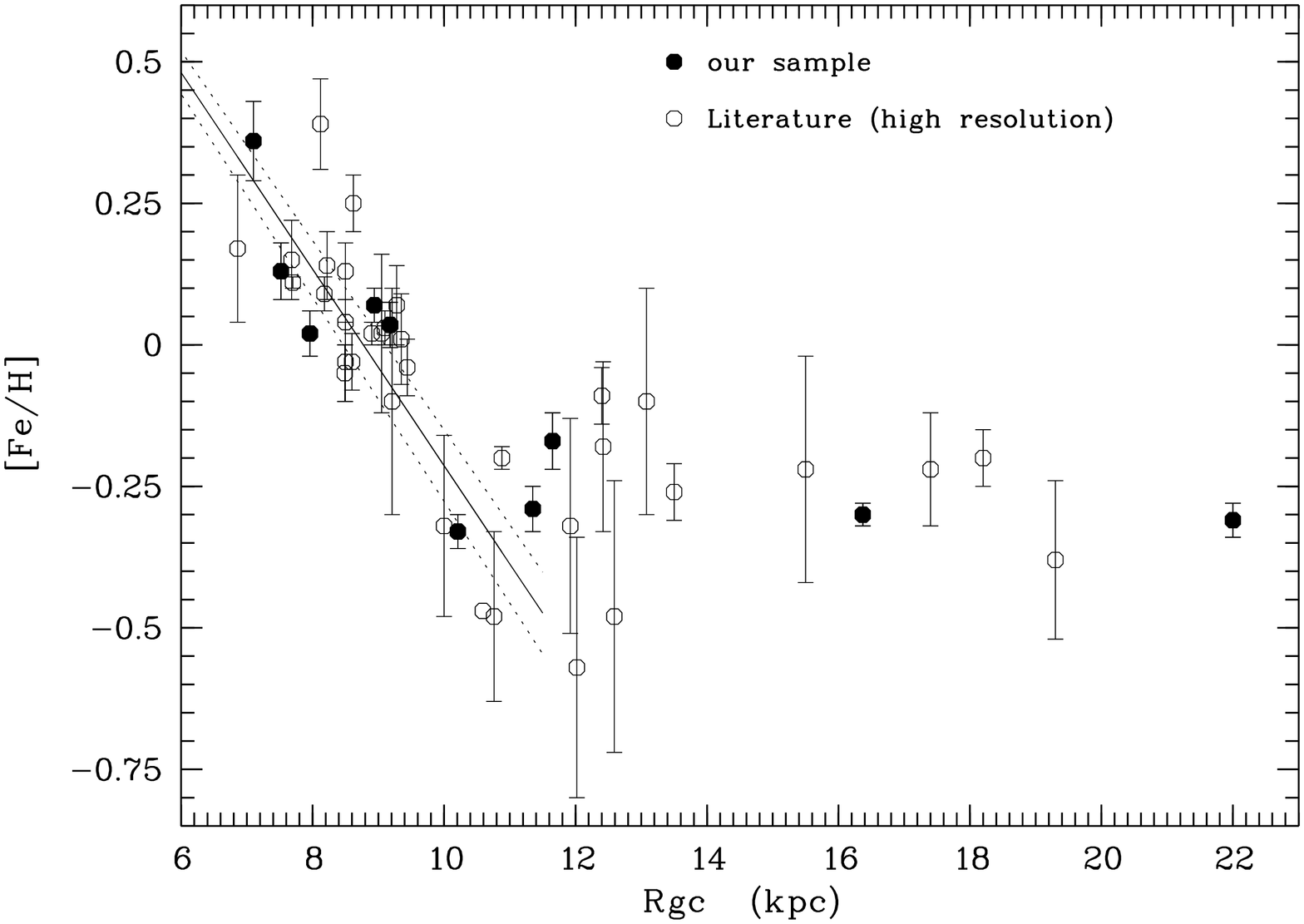, width=6cm,
angle=-90}\psfig{figure=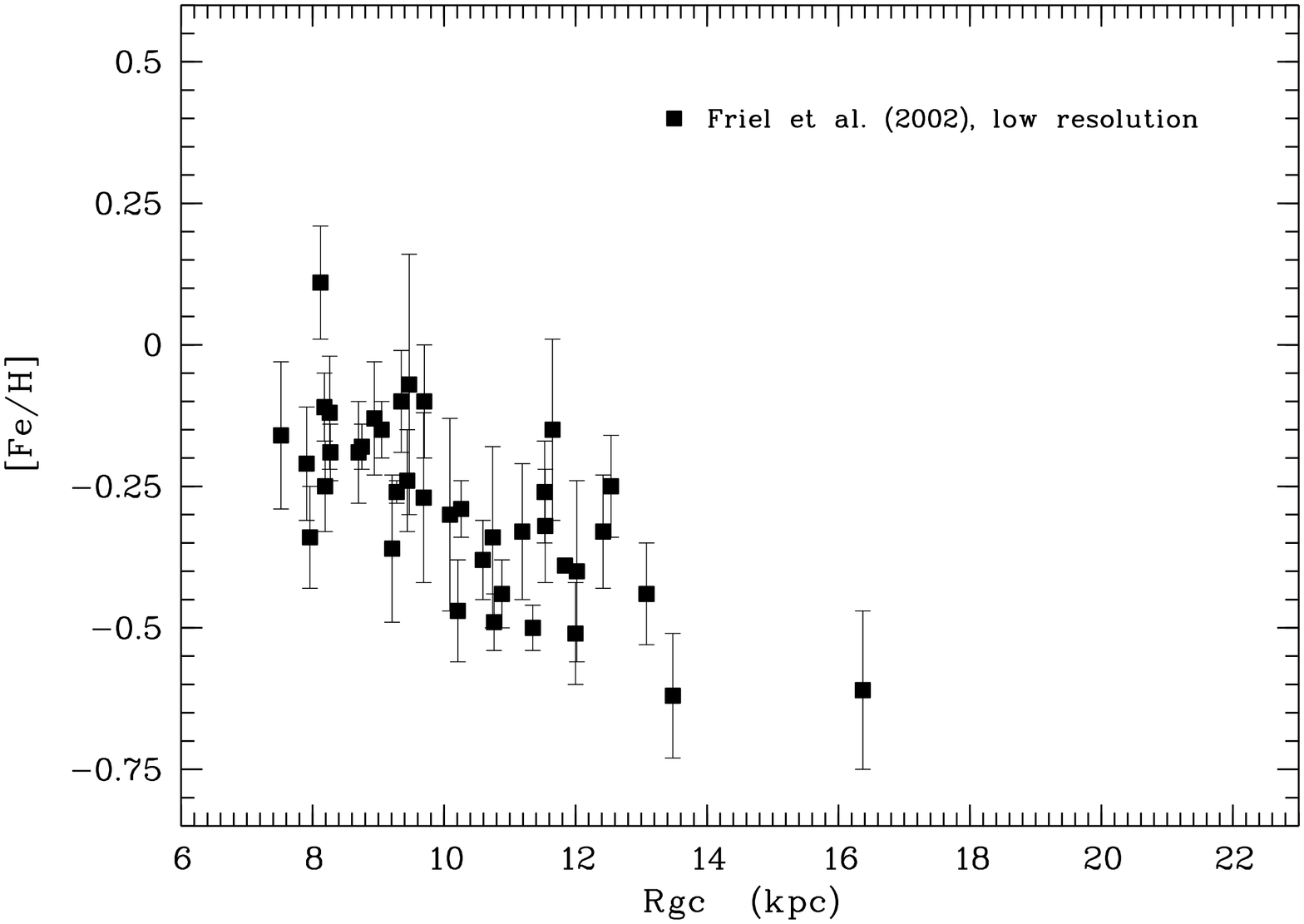, width=6cm, angle=-90} \caption{Radial
gradient ([Fe/H] vs.~Galactocentric distance) for open clusters. Left panel:
the results for clusters in our sample analyzed so far (filled circles; this
paper, Papers~{\sc i}, {\sc ii}, and S07) are compared to other clusters
analyzed with high-resolution spectroscopy (open circles). Right panel: the
low resolution sample by Friel et al. (2002).}\label{gradiente}
\end{figure*}

\begin{figure*}
\psfig{figure=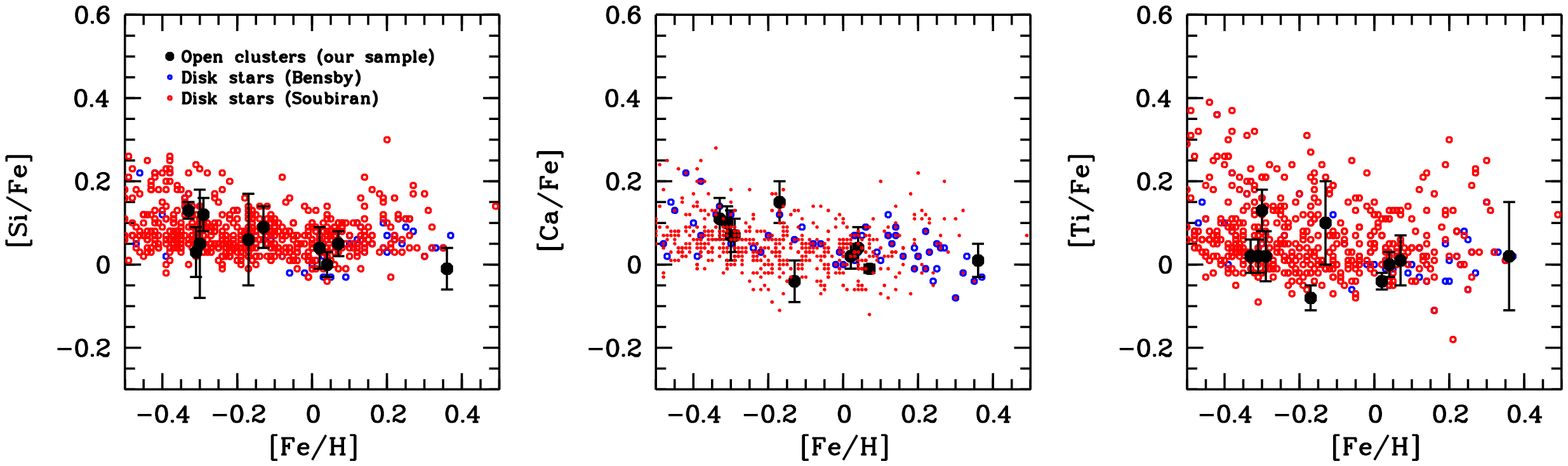, width=7cm, angle=-90}  %

\psfig{figure=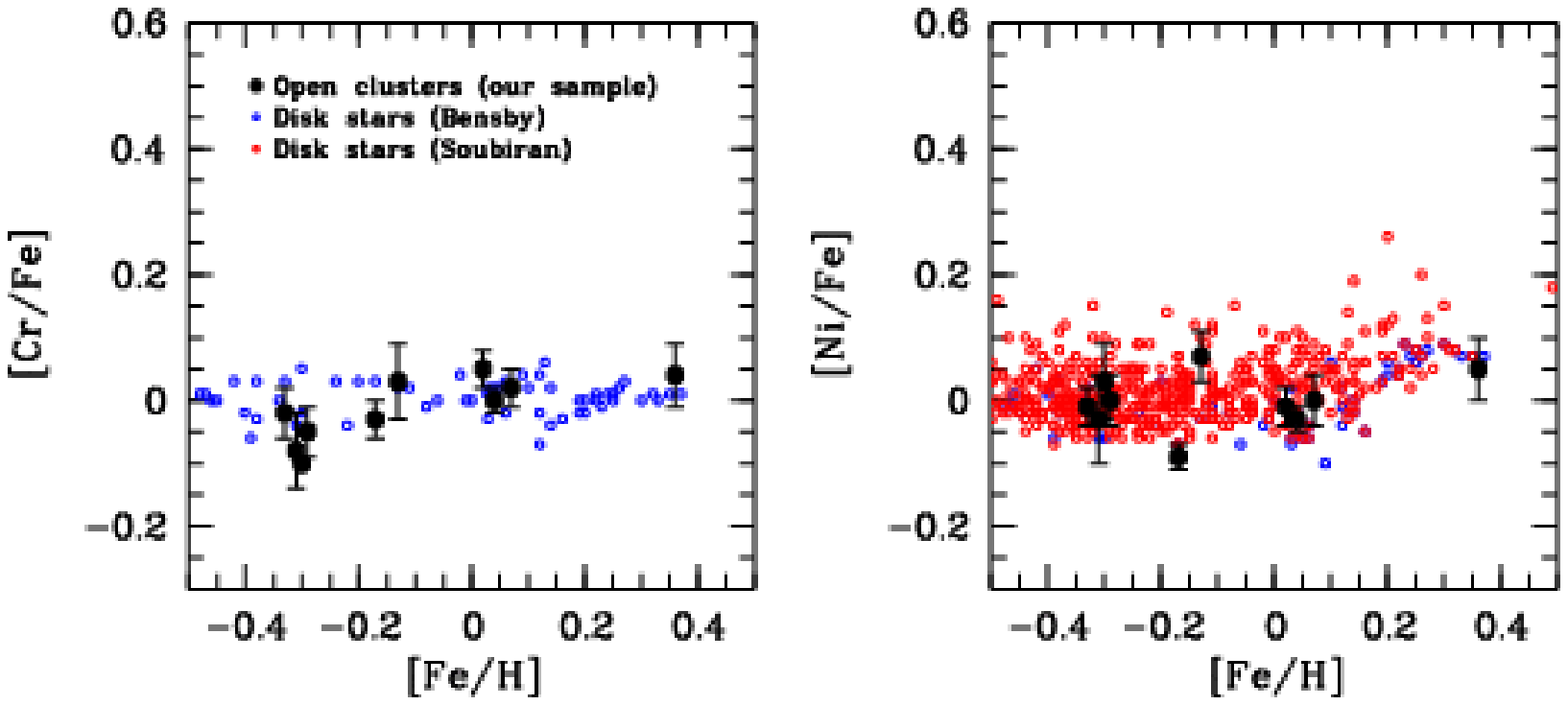, width=7cm, angle=-90} %

\caption{[X/Fe] vs. [Fe/H]: comparison between open clusters in our sample
(black filled circles) and disk stars (dots) by Bensby et al. (\cite{bensby},
blue) and Soubiran \& Girard (\cite{soubiran}, red).}\label{confronto_disco}
\end{figure*}

\begin{figure*}
\psfig{figure=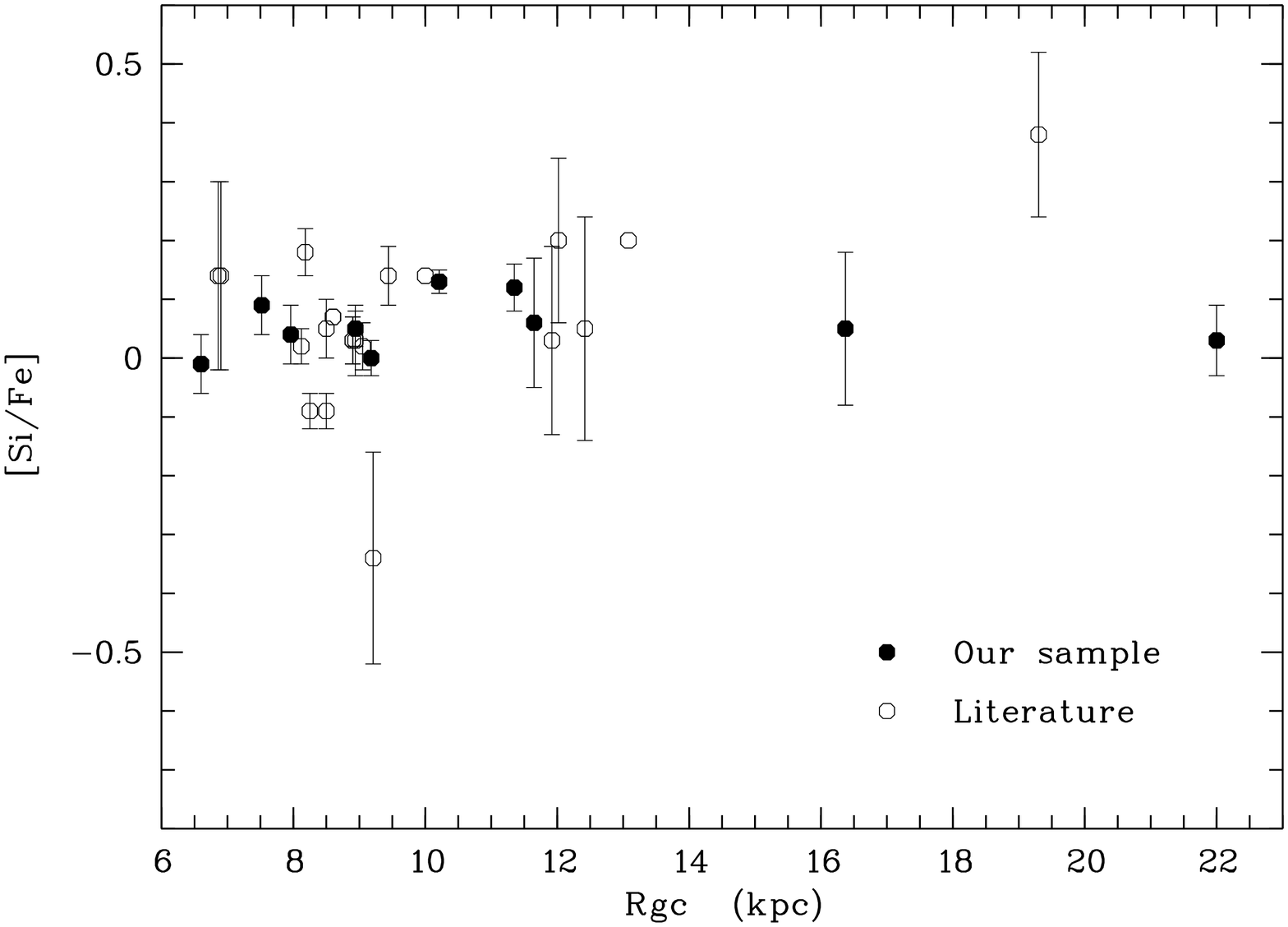, width=6cm, angle=-90}
\psfig{figure=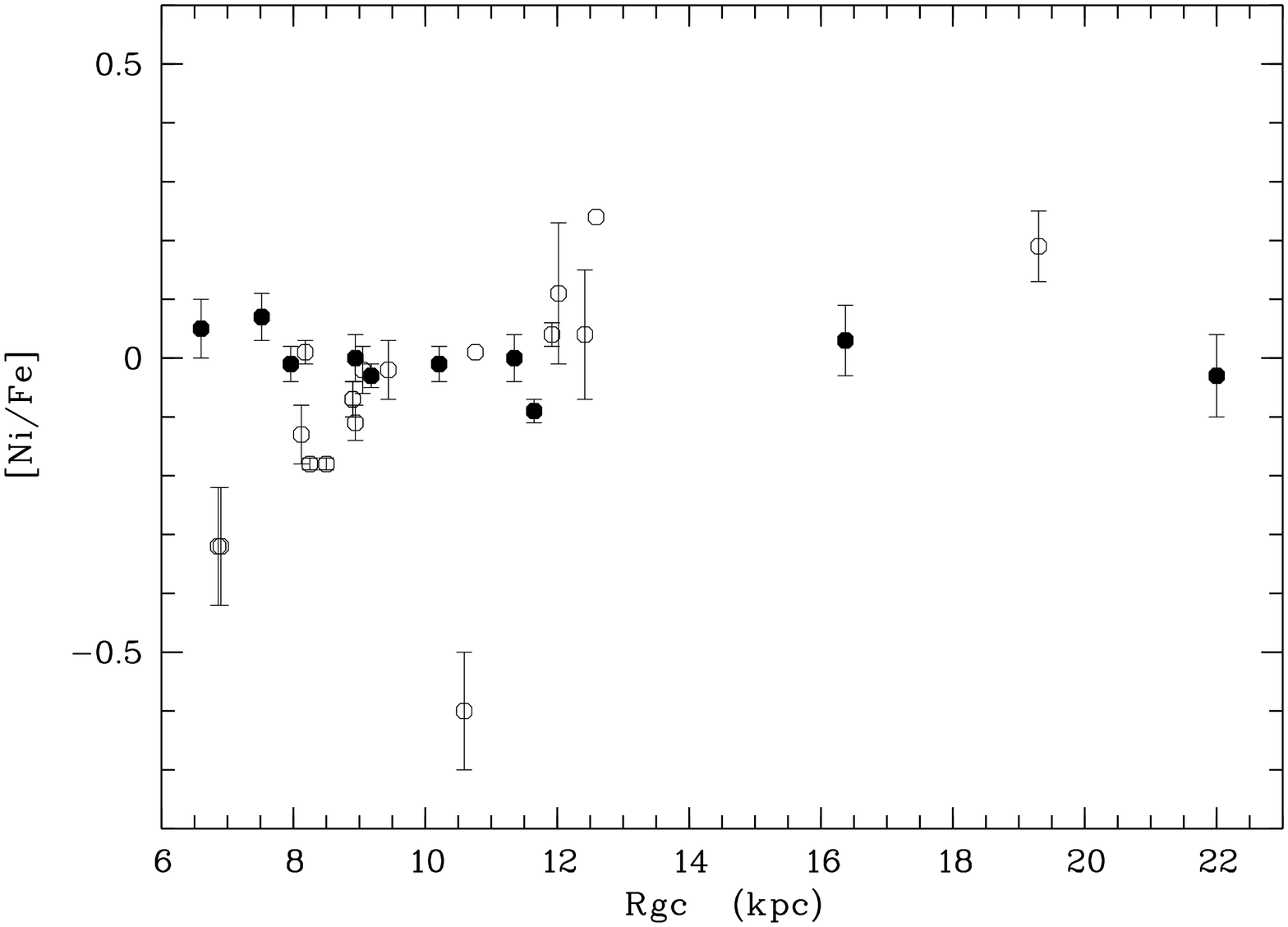, width=6cm, angle=-90}

\psfig{figure=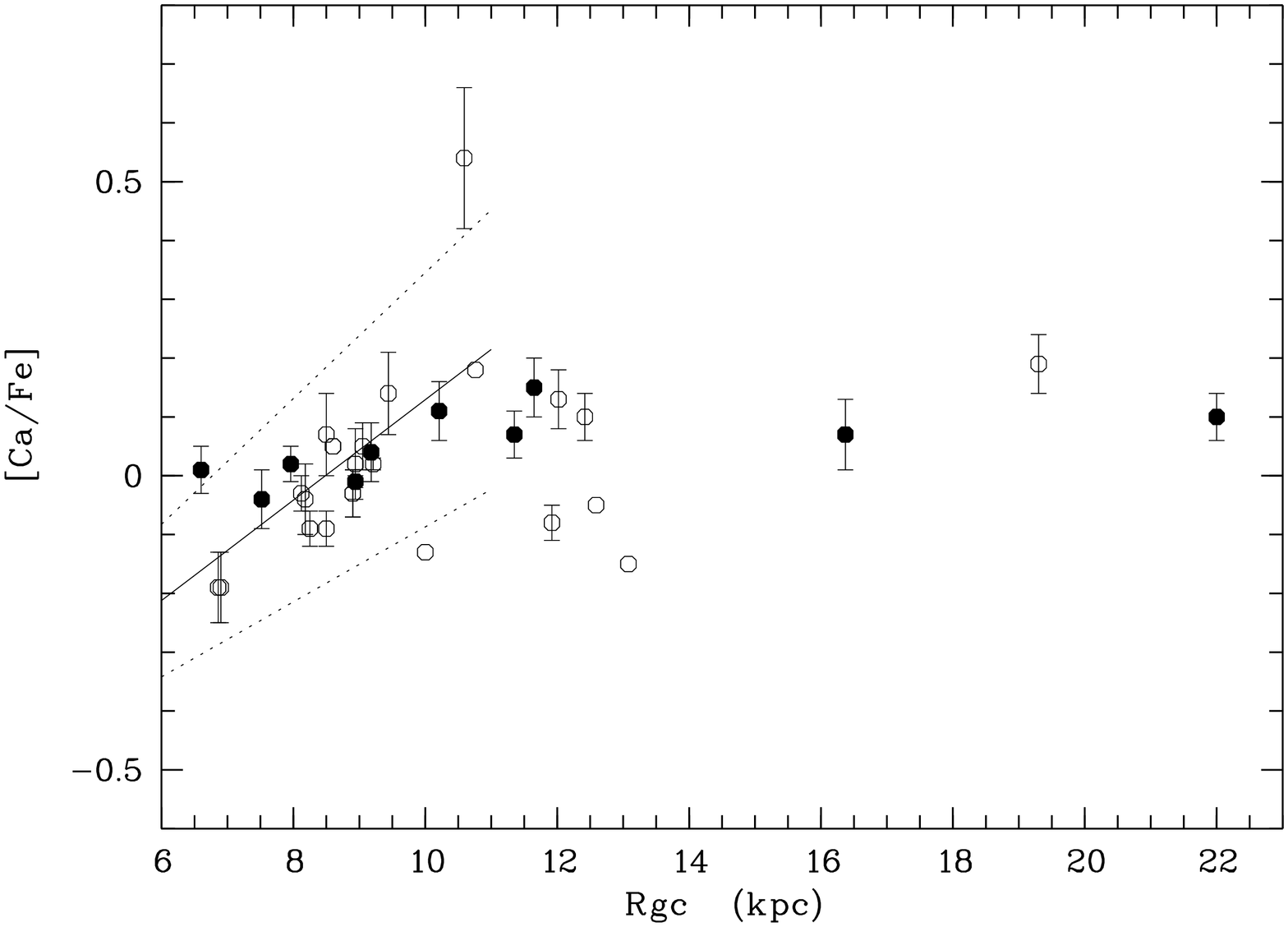, width=6cm, angle=-90}
\psfig{figure=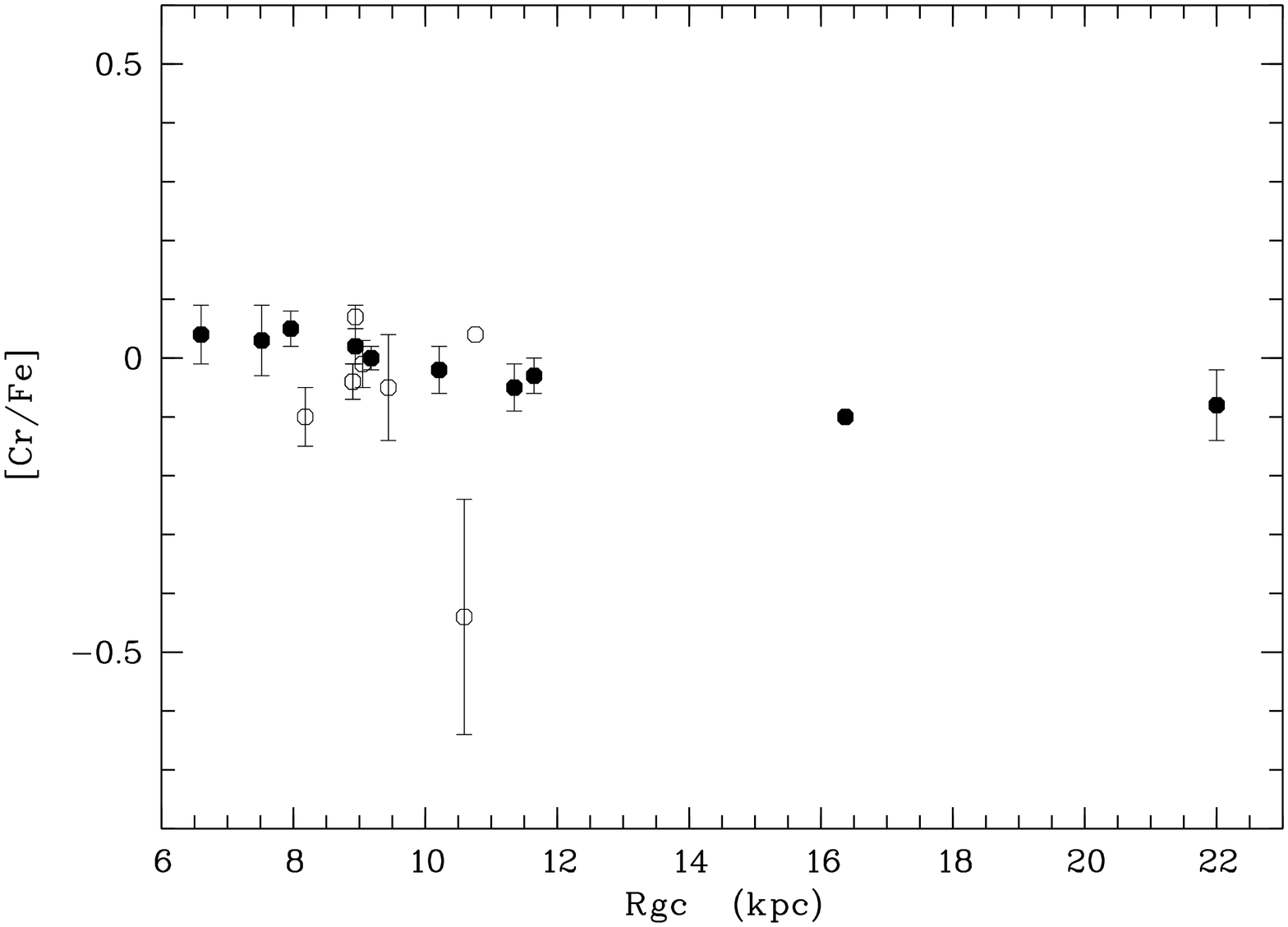, width=6cm, angle=-90}

\psfig{figure=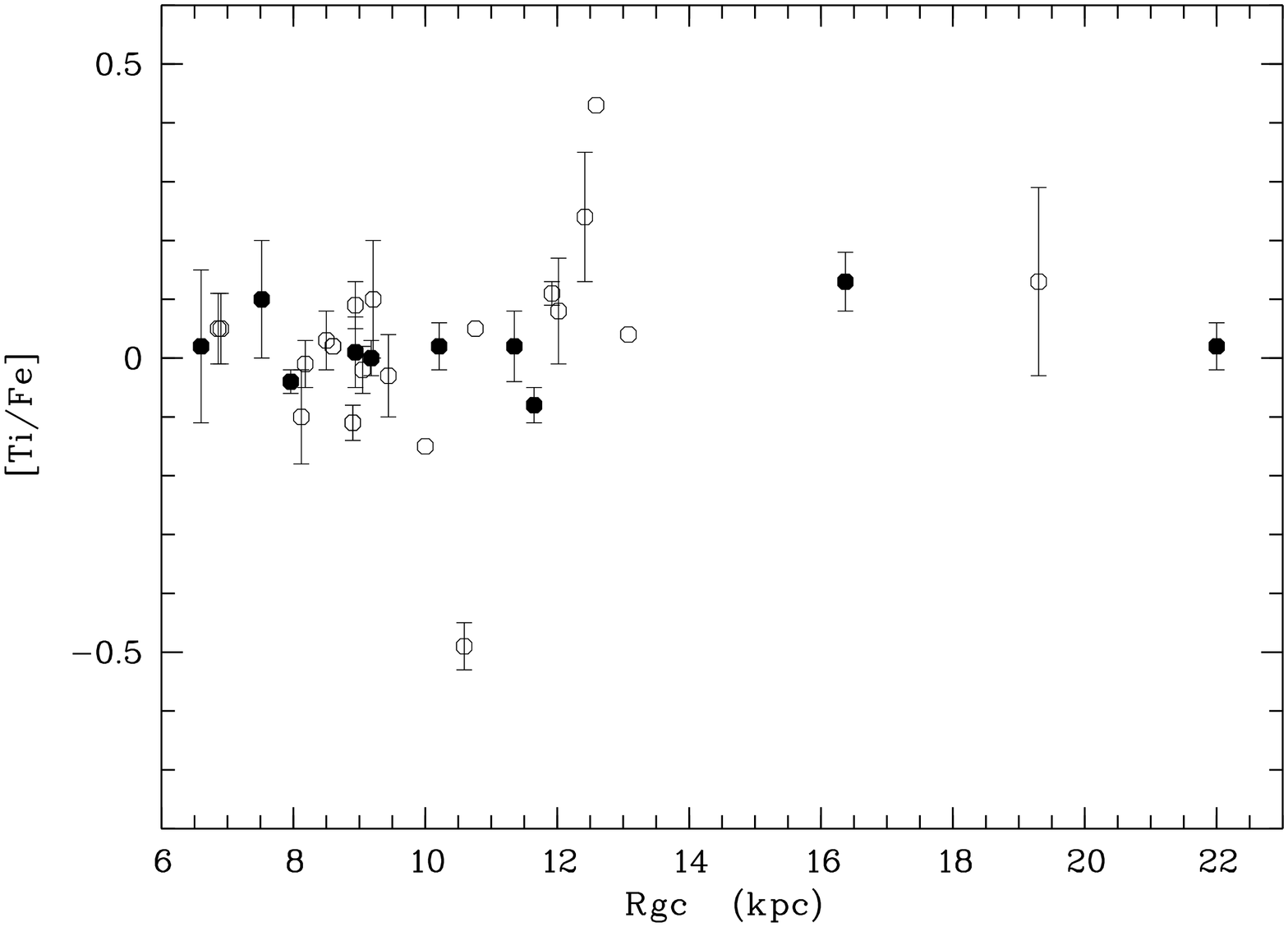, width=6cm, angle=-90}
\psfig{figure=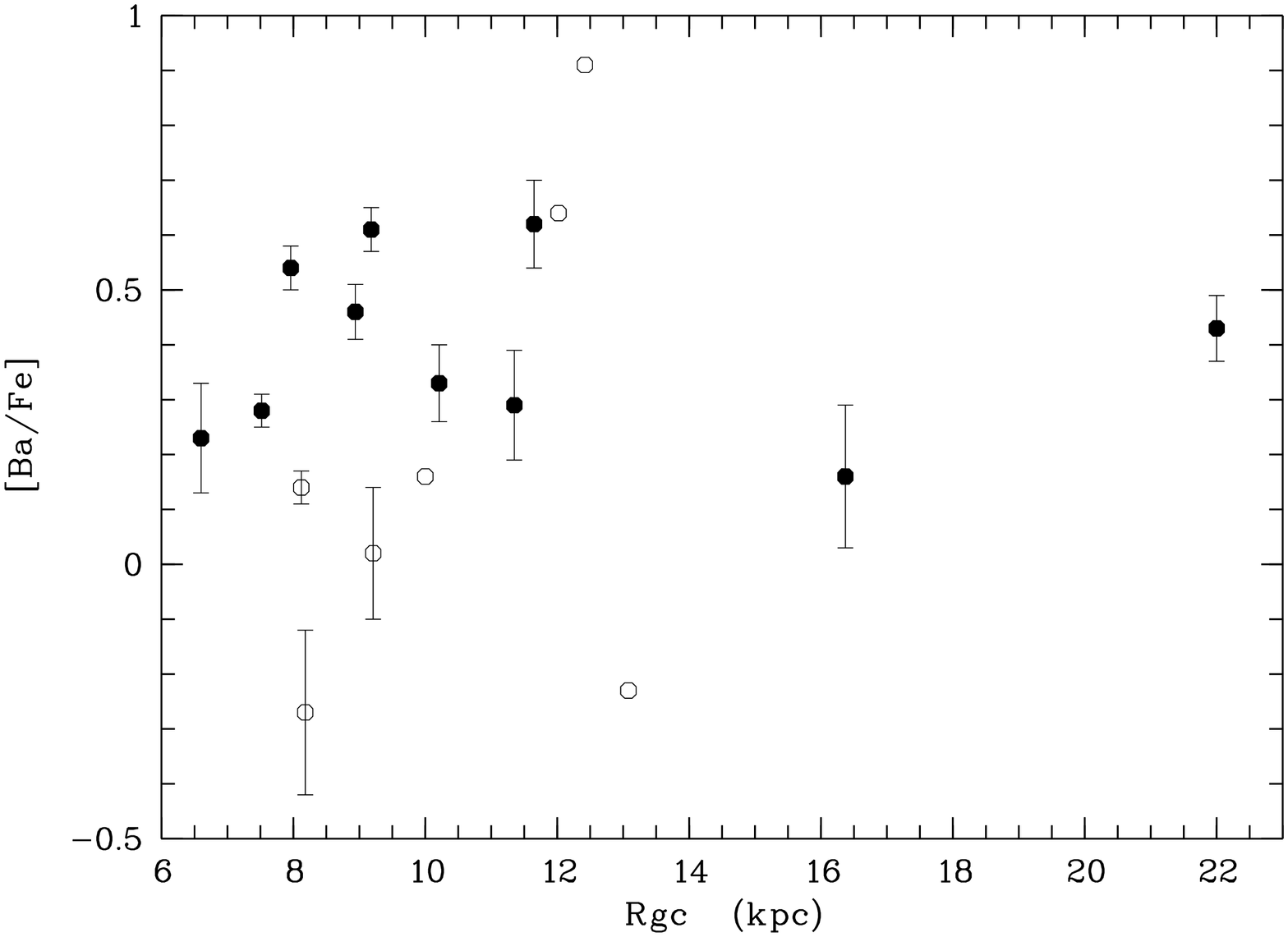, width=6cm, angle=-90}

\caption{Radial gradients of Si, Ca, Ti, Cr, Ni, and Ba, based on high
resolution data: our sample (filled circles), and other literature
investigations (open circles).}\label{grad_altri}
\end{figure*}

\end{document}